\documentclass[mnsc,nonblindrev]{informs_modified} 

 \OneAndAHalfSpacedXII

\usepackage{endnotes}
\let\footnote=\endnote
\let\enotesize=\normalsize
\def\notesname{Endnotes}%
\def\makeenmark{$^{\theenmark}$}
\def\enoteformat{\rightskip0pt\leftskip0pt\parindent=1.75em
	\leavevmode\llap{\theenmark.\enskip}}

\usepackage{natbib}
\bibpunct[, ]{(}{)}{,}{a}{}{,}%
\def\bibfont{\small}%
\TheoremsNumberedThrough     
\ECRepeatTheorems

\EquationsNumberedThrough    

\usepackage{setspace}
\usepackage{amsmath}
\usepackage{amssymb}  
\usepackage{amsfonts}
\usepackage{array}
\usepackage{booktabs}
\usepackage{comment}
\usepackage{graphicx}
\usepackage{multirow} 
\usepackage{mathtools}
\usepackage{subfigure}
\usepackage{tabu} 
\usepackage{tabularx}
\usepackage{tabulary}
\usepackage[colorinlistoftodos]{todonotes}
\definecolor{DarkBlue}{rgb}{0.0,0.0,0.55}
\usepackage[backref = false, bookmarks, colorlinks = true, plainpages = false, citecolor = DarkBlue , urlcolor = DarkBlue, filecolor = DarkBlue, linkcolor = DarkBlue]{hyperref}
\usepackage{cleveref}
\usepackage{xurl}
\usepackage{pifont}
\usepackage{bm}
\usepackage{threeparttable}

\usepackage{geometry}
\usepackage{amsmath,amsfonts,bm}

\def\eqref#1{equation~\ref{#1}}
\def\Eqref#1{Eq.~(\ref{#1})}

\def\1{\bm{1}}

\def\Path{{\triangle\mathcal{P}^{(t,c)}}}
\def\path{{\mathcal{P}}}

\def\vw{{\bm{w}}}

\def\mP{{\bm{P}}}

\def\mW{{\bm{W}}}

\DeclareMathAlphabet{\mathsfit}{\encodingdefault}{\sfdefault}{m}{sl}
\SetMathAlphabet{\mathsfit}{bold}{\encodingdefault}{\sfdefault}{bx}{n}

\def\gA{{\mathcal{A}}}

\def\gC{{\mathcal{C}}}
\def\gD{{\mathcal{D}}}
\def\gE{{\mathcal{E}}}
\def\gF{{\mathcal{F}}}
\def\gG{{\mathcal{G}}}

\def\gM{{\mathcal{M}}}
\def\gN{{\mathcal{N}}}

\def\gP{{\mathcal{P}}}

\def\gS{{\mathcal{S}}}

\def\gV{{\mathcal{V}}}

\def\sD{{\mathbb{D}}}

\def\sI{{\mathbb{I}}}

\def\sM{{\mathbb{M}}}

\def\sS{{\mathbb{S}}}

 \newcommand{\E}{\mathbb{E}}

 \newcommand{\prob}{\mathbb{P}}

\newcommand{\Var}{\mathrm{Var}}

\newcommand{\Cov}{\mathrm{Cov}}
\newcommand{\Corr}{\mathrm{Corr}}
\newcommand{\Diff}{{\triangle}\sM^{(t,c)}}
\newcommand{\ATE}{\tau}
\newcommand{\Naive}{\hat{\tau}_{\text{Naive}}}
\newcommand{\AP}{\hat{\ATE}_{\text{AP}}}
\newcommand{\EPTE}{\hat{\Gamma}}
\newcommand{\PTE}{\Gamma}
\newcommand{\inde}{\text{deg}^-}
\newcommand{\outde}{\text{deg}^+}

\usepackage{algorithmic}
\usepackage[ruled,linesnumbered]{algorithm2e}

\begin{document}
	
	
	\RUNAUTHOR{Wang}
	
	\RUNTITLE{Test of Matching}
	
	\TITLE{Experimental Design for Matching}
	
	\ARTICLEAUTHORS{%
	\AUTHOR{Chonghuan Wang}
	\AFF{Naveen Jindal School of Management, University of Texas at Dallas, Richardson, TX 75080, \EMAIL{chonghuan.wang@utdallas.edu}} 
	} 
	
 	\ABSTRACT{Matching mechanisms play a central role in operations management across diverse fields including education, healthcare, and online platforms. However, experimentally comparing a new matching algorithm against a status quo presents some fundamental challenges due to matching interference, where assigning a unit in one matching may preclude its assignment in the other. In this work, we take a \textit{design-based perspective} to study the design of randomized experiments to compare two predetermined matching plans on a finite population, without imposing outcome or behavioral models. We introduce the notation of a disagreement set, which captures the difference between the two matching plans, and show that it admits a unique decomposition into disjoint alternating paths and cycles with useful structural properties. Based on these properties, we propose the Alternating Path Randomized Design, which sequentially randomizes along these paths and cycles to effectively manage interference. Within a minimax framework, we optimize the conditional randomization probability and show that, for long paths, the optimal choice converges to $\sqrt{2}-1$, minimizing worst-case variance. We establish the unbiasedness of the Horvitz-Thompson estimator and derive a finite-population Central Limit Theorem that accommodates complex and unstable path and cycle structures as the population grows. Furthermore, we extend the design to many-to-one matchings, where capacity constraints fundamentally alter the structure of the disagreement set. Using graph-theoretic tools, including finding augmenting paths and Euler-tour decomposition on an auxiliary unbalanced directed graph, we construct feasible alternating path and cycle decompositions that allow the design and inference results to carry over. 
    }
	
	
	
	\maketitle

\section{Introduction}
Matching is a foundational topic in operations management, with deep theoretical roots and wide-ranging real-world deployment. Across domains such as school choices and admission (e.g., \citealt{chen2021information,ignacio2025stable,gong2025dynamic}), organ exchange and transplantation (e.g., \citealt{roth2004kidney,ashlagi2021kidney, tang2025multi}), refugee settlement (e.g., \citealt{delacretaz2023matching,ahani2021placement}), ridesharing platforms (e.g., \citealt{freund2024two}), dating markets (e.g., \citealt{hitsch2010matching, rios2023improving}) and video game (e.g., \citealt{chen2021matchmaking}), matching mechanisms serve as critical infrastructure for allocating scarce resources, forming partnerships, and ultimately increasing social welfare. The resulting economic and social impact has made matching a central topic of interest to both researchers and practitioners.

A standard approach in both research and practice is to model agent behavior or forecast the matching utilities, and then design mechanisms that perform well under those models. While this modeling approach has yielded important insights, practitioners may still worry about \textit{the reliability of such assumptions} when deploying mechanisms in real-world systems. This naturally raises the question of how one can empirically evaluate the performance of a newly proposed matching mechanism. Moreoever, as digitization increases the granularity of observable features used in matching, it becomes common to ask whether incorporating additional features actually improves matching quality, and whether such improvements justify the associated engineering and financial costs. These considerations motivate a complementary perspective focused on \textit{empirical testing} rather than \textit{structural modeling}. In this work, we study a basic but important question: \textbf{How can we experimentally compare two given matching plans on the same finite population without imposing outcome or behavioral models?}

More concretely, we consider a setting in which two predetermined matching plans can be applied to the same finite population: a \textit{treatment matching}, generated by a new mechanism or using additional features, and a \textit{control matching}, representing the status quo. We define the treatment effect as the difference in average outcomes under the treatment and control matchings. Our objective is to design valid randomized experiments that allow us to estimate and infer this treatment effect. We adopt a \textit{design-based perspective}, meaning that inference relies on the known randomization scheme rather than a model of matching outcomes or agent behavior. This enables us to draw conclusions without imposing outcome or behavioral models, which is essential when matching assumptions themselves are under investigation.

The core technical challenge arises because a unit may be paired with different counterparts under the treatment and control matchings. As a result, in an experiment we can only observe the outcome of one realized match, but never the counterfactual outcome of the alternative match. This lack of individual-level counterfactuals is a familiar obstacle in standard causal inference. Beyond this, the structure of matching introduces \textit{hard feasibility} constraints that couple the randomization decisions across units. In classical experimental designs, assigning one unit to treatment or control is typically independent of assignments for other units. In contrast, when deciding whether to adopt a match from the treatment or control plan, that decision may preclude other feasible matches due to capacity or compatibility constraints, and the resulting impact can propagate across the matching. We refer to this phenomenon as matching interference. Because matching pairs cannot be treated as independent experimental units, a central task in our setting is to design randomization schemes that carefully manage and quantify such interference to enable valid estimation and inference.

\subsection{Main Results}

We start with arguably the fundamental case of designing randomized experiments for one-to-one matchings. We introduce the notion of the \textit{disagreement set}, defined as the set of matching pairs that appear in exactly one of the two matchings (treatment or control). This set fully characterizes the difference between the two matchings and therefore contains all pairs that require randomization. A key structural observation is that the disagreement set admits a \textit{unique decomposition} into disjoint paths and cycles, where edges correspond to matching pairs and alternate between the treatment matching and the control matching. Along any such path or cycle, adjacent edges cannot be simultaneously realized in the experiment due to capacity constraints. This motivates our \textit{Alternating Path Randomized Design} (AP design), in which randomization proceeds sequentially along each path or cycle to ensure feasibility. Intuitively, the decision to include an edge is randomized conditional on the realized selection of its predecessor along the path or cycle.


We next optimize the conditional randomization probabilities of our design using a \textit{minimax} framework. We show that, in the worst case and as paths and cycles grow long, the optimal conditional probability converges to $\sqrt{2}-1$, which is strictly less than $0.5$. Intuitively, this reflects a trade-off: reducing the variance contribution of the current edge while accounting for downstream edges along the path or cycle. For inference, we establish a finite-population central limit theorem (CLT) for the Horvitz–Thompson estimator, enabling valid asymptotic inference under the design-based perspective. To our knowledge, this result is not a direct consequence of existing CLTs in the literature. The technical novelty arises from the fact that the path/cycle decomposition can exhibit highly heterogeneous and potentially unstable patterns: the number of components may be large with short lengths, small with extremely long lengths, or fluctuate between these regimes without converging to a stable structure as the population grows. Existing tools typically target either many short components or a few long ones, but not (unstable) mixtures that change with the size of the population. We address this by combining classical Lindeberg–Feller arguments with $\alpha$-mixing techniques through Bolzeno-Weierstrass theorem and the subsequence principle, yielding a CLT that accommodates such instability.


We further consider the experimental design problem for many-to-one matchings. In this setting, a new and significant challenge arises: the path/cycle decomposition of the disagreement set is generally \textit{not unique}, and many possible decompositions violate capacity constraints and are therefore not admissible for randomization. A key part of our contribution is to identify \textit{valid decompositions} that preserve feasibility so that our AP design and its inference guarantees remain applicable. We provide a set of \textit{sufficient conditions} under which a decomposition is admissible, and show that finding such a decomposition can be reduced to two graph-theoretic problems, \textit{finding augmenting paths} and \textit{Euler-tour decompositions}, via the construction of an auxiliary unbalanced directed graph and a flow network. This extension from one-to-one to many-to-one matchings is structurally non-trivial, as it requires additional combinatorial feasibility considerations beyond those present in the one-to-one case.


    \section{Related Work}

Matching, as one of the most common practice in today's platform operations, has attracted much attention in running experiments on platforms. The main focus of the current literature is that the matching introduces interference among experimental units (see, e.g., \citealt{johari2022experimental}). More specifically, individuals in the market compete with each other through a matching process. \cite{li2022interference} and \cite{dhaouadi2023price} develop insightful market models to study bias and variance in two-sided markets with interference. \cite{weng2024experimental} carefully investigate the experimentation on one-side markets, such as the live-interaction platforms and the video games. To mitigate the bias, \cite{bright2025reducing} consider a matching mechanism based on linear programming and debias the estimators via shadow prices. \cite{bajari2023experimental} and \cite{masoero2024multiple} propose the two-sided randomization to reduce the bias, which are also known as multiple randomization designs. \cite{wager2021experimenting} take an equilibrium model into account and study the local experimentation of the two-sided market. \cite{pouget2019variance}, \cite{doudchenko2020causal}, \cite{harshaw2023design} introduce the framework of bipartite experiments. \cite{zhan2024estimating} and \cite{zhang2025debiasing}  investigate seller-side experiments by using the MNL model to capture the interference introduced by the two-sided market. If the market has some Markovian properties, \cite{farias2022markovian} propose a novel
model-free difference-in-Q estimator and \cite{chen2024experimenting} showcase how the treatment locality can help to further reduce the variance. All the mentioned works try to mitigate the bias issue introduced by a fixed matching mechanism. However, they do not discuss how to test or compare two given matching mechanisms, which is the scope of this paper.

More broadly, the paper is related to the growing literature of designing experiments in complex systems with interference. One of the most well-studied interferences is network effect (see, e.g., \citealt{eckles2017design,athey2018exact,li2022random,ugander2023randomized,viviano2023causal,cortez2023exploiting, candogan2024correlated,holtz2025reducing}). A large stream of remedy in the recent literature is to run switchback experiment (see, \citealt{bojinov2023design}, \citealt{hu2022switchback}, \citealt{xiong2024data}, \citealt{ni2025enhancing}, \citealt{chen2025efficient}, \citealt{ni2025reliable}). In a switchback experiment, the entire platform
alternates between treatment and control usually on an hourly basis. In addition, recent academic focus has shifted to link the decision-making to experimental design, e.g., \cite{dhaouadi2023price}, \cite{johari2024does}, \cite{li2025choosing} and \cite{ni2025decision}. \cite{johari2024does} reveal the scenarios where the interference does not matter in terms of decision-making, although it biases the estimation. In contrast, \cite{li2025choosing} identify that when testing recommendation algorithms, the interference may alter the sign of the treatment effects leading to wrong decision of deploying the new algorithms or not. \cite{ni2025decision} adopts a robust optimization approach to suggest the optimal decision rule on adopting the new treatment based on the observations from switchback experiments. For experimentation in operational systems, \cite{li2023experimenting} investigate the experimentation under queuing congestion, and \cite{chen2025bias} study the bias of experiment in complex inventory problem. \cite{xiong2025automated} recently introduces the simulation tool to tackle the challenge that the objective is typically a black-box function of the design.  Also, much work centers on identifying individualized treatment rules to maximize the welfare, i.e., policy learning (\citealt{kitagawa2018should}, \citealt{kallus2021minimax}, \citealt{zhang2024minimax}). We refer to \cite{wager2024causal} for a detailed discussion on policy learning. In addition, causal message passing by \cite{shirani2024causal} can be powerful to model very general network interference in experimentation.


\section{Preliminaries}
\subsection{Model}
We begin with the \textit{one-to-one} matching setting for clarity. Consider a population of $2N$ agents that need to be matched in pairs. Our formulation accommodates both one-sided environments (e.g., pairing two players in a one-versus-one game) and two-sided environments (e.g., matching $N$ workers to $N$ jobs). In the two-sided case, we assume for simplicity that the population consists of $N$ agents on each side. A matching is represented by a function
\begin{equation*}
    m: [2N]\rightarrow [2N]\cup \{0\},
\end{equation*}
where $[2N]$ represents the set $\{1,2,\cdots,2N\}$, $m(i)$ denotes the agent matched to $i$, and $m(i)=0$ indicates that $i$ is unmatched. A one-to-one matching is feasible if $m(i)=j>0$ guarantees $m(j)=i$ for any agent $i$, so that matches are symmetric and an agent cannot be matched to multiple partners. We let $\sM$ denote the set of all the realized match pairs:
\begin{equation}\notag
    \sM=\{(i,m(i)) : 1\le i < m(i) \le 2N \},
\end{equation}
where the condition $i<m(i)$ ensures that each pair appears exactly once and precludes the unmatched agents, i.e., the agent with $m(i)=0$.


\begin{figure}[t]
    \centering
    \includegraphics[width=0.92\linewidth]{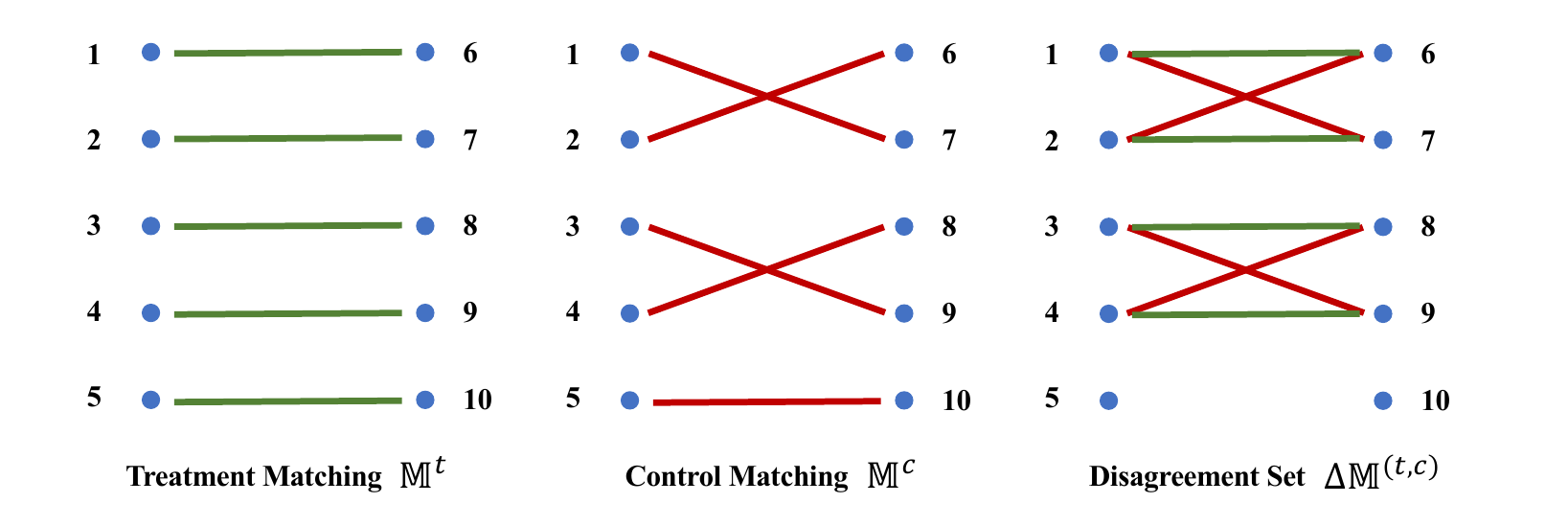}
    \caption{An illustrating example of $\sM^t$, $\sM^c$ and the disagreement set $\Diff$.}
    \label{fig:matching-example}
\end{figure}

We are given two matchings on the same set of $2N$ agents: a treatment matching $\sM^t$ and a control matching $\sM^c$. The treatment matching $\sM^t$ can be proposed by a new matching algorithm considering new features and the control matching $\sM^c$ is from the status quo. For an agent $i$, we denote its partner in $\sM^t$ by $m^t(i)$ and in $\sM^c$ by $m^c(i)$. Because each agent can be matched with at most one partner, an agent cannot simultaneously realize both $m^t(i)$ and $m^c(i)$, which motivates the need to compare the two matchings through an experiment rather than direct observation. To formalize the difference between $\sM^t$ and $\sM^c$, we define the disagreement set
\begin{equation}\label{eq:diff-1-1}
    \Diff = (\sM^t \cup \sM^c) \backslash (\sM^t \cap \sM^c),
\end{equation}
which contains all match pairs that appear in exactly one of the two matchings. We further partition $\Diff$ into
\begin{equation*}
    \Diff_t = \Diff \cap \sM^t \text{ and } \Diff_c = \Diff \cap \sM^c,
\end{equation*}
which we refer to respectively as \textit{t-matches} and \textit{c-matches}. Thus, $\Diff_t$ collects the match pairs that would be realized only under the treatment matching, and $\Diff_c$ collects those realized only under the control matching. Comparing these pairs is the core object of the experimental design.

\begin{example}
Figure~\ref{fig:matching-example} illustrates a toy instance with $2N=10$ agents and two one-to-one matchings. The pair $(5,10)$ appears in both $\sM^t$ and $\sM^c$ and is therefore excluded from the disagreement set. The resulting disagreement set is
\[
\Diff = \{(1,6),(1,7),(2,6),(2,7),(3,8),(3,9),(4,8),(4,9)\}.
\]
Among these, the t-matches are $\Diff_t = \{(1,6),(2,7),(3,8),(4,9)\}$, and the c-matches are $\Diff_c = \{(1,7),(2,6),(3,9),(4,8)\}$.
\end{example}

We associate each ordered pair $(i,j)$ with a \textit{potential outcome} $Y_{i,j}$, representing the outcome that would be realized if $i$ were matched with $j$.  We make no structural assumptions or behavior models on $Y_{i,j}$. Under the design-based perspective (e.g., \citealt{imbens2015causal}), these values are treated as fixed but unknown, and randomness arises solely from the experimental design. The outcome $Y_{i,j}$ is revealed only if the match $(i,j)$ is realized in the experiment. Our estimand of interest is the average treatment effect (ATE) of the treatment matching $\sM^t$ relative to the control matching $\sM^c$:
\begin{equation*}
    \ATE= \bar{Y}_t-\bar{Y}_c,\quad \bar{Y}_t= \frac{\sum_{(i,j)\in \sM^t} Y_{i,j}}{N}, \quad \bar{Y}_c=\frac{\sum_{(i,j)\in \sM^c} Y_{i,j}}{N}.
\end{equation*}
If $m^t(i)\not= m^c(j)$, then at most one of $Y_{i,m^t(i)}$ and $Y_{i,m^c(i)}$ can be observed, so $\ATE$ cannot be directly observed and must be estimated from an experiment. Since match pairs appearing in both $\sM^t$ and $\sM^c$ contribute equally to $\bar{Y}_t$ and $\bar{Y}_c$ and cancels out in $\tau$, the disagreement set $\Diff$ contains exactly the match pairs relevant for identification.

Given these definitions, an experimental design for matching consists of two elements: (i) a randomization scheme that selects a feasible set of realized matches $\sM^{\text{EXP}}$ in the presence of matching interference, and (ii) an estimator $\hat{\ATE}$ that uses the realized outcomes and the randomization process to estimate $\ATE$.

Before introducing our proposed design, it is instructive to consider a simple baseline that highlights the limitations of directly switching between the two matchings.

\begin{remark}[The Naive Experimental Design] \label{remark:naive}
A natural baseline design selects either the treatment matching $\sM^t$ or the control matching $\sM^c$ with equal probability $1/2$. Let $\sM^{\mathrm{Naive}}$ denote the realized matching and define 
\[
\bar{Y}^{\mathrm{Naive}} = \frac{1}{N} \sum_{(i,j)\in \sM^{\mathrm{Naive}}} Y_{i,j}.
\]
Under this design, $\bar{Y}^{\mathrm{Naive}} = \bar{Y}_t$ with probability $1/2$ and $\bar{Y}^{\mathrm{Naive}} = \bar{Y}_c$ with probability $1/2$. A natural inverse-propensity estimator of $\ATE$ is
\[
\hat{\tau}_{\mathrm{Naive}}
= 2\,\bar{Y}^{\mathrm{Naive}}\left(
\mathbb{I}\{\sM^{\mathrm{Naive}}=\sM^t\}
-
\mathbb{I}\{\sM^{\mathrm{Naive}}=\sM^c\}
\right).
\]


    \begin{proposition}[Bias and Variance of the Naive Design] \label{prop:naive-design} Under the naive design, the estimator is unbiased $\E(\Naive)=\tau$ with the variance $\Var(\Naive)=(\bar{Y}_t+\bar{Y}_c)^2$.
    \end{proposition}
    
    Although unbiased, the estimator exhibits two drawbacks. First, its variance does not vanish as $N\to\infty$ in general. For example, in a two-sided market with $N$ workers and $N$ jobs, suppose each realized match yields an outcome $Y_{i,j}\ge c_0>0$. Then for any two complete matchings, $\mathrm{Var}\!\left(\hat{\tau}_{\mathrm{Naive}}\right)\ge c_0^2$ for any $N>0$, so increasing the experimental population does not improve precision. Second, $\widehat{\tau}^{\mathrm{Naive}}$ takes only two possible values and therefore does not satisfy a CLT that would enable asymptotic inference.
\end{remark}
    


\section{Alternating Path Randomized Design}
We now introduce our proposed Alternating Path Randomized Design (AP Design) for comparing two predetermined matchings in the presence of matching interference. The key structural insight underlying AP design is that the disagreement set $\Diff$ can be decomposed into several alternating paths or cycles, which are defined as follows.
\begin{definition}[Alternating Path/Cycle]\label{def:alternating-path}
    An alternating path (or alternating cycle) in $\Diff$ is an ordered sequence of agents $\{v_1,v_2\cdots,v_{k+1}\}$ such that:
    \begin{itemize}
        \item For every $i \in \{1,\dots,k\}$, the edge $(v_i, v_{i+1})$ belongs to $\Diff$.
        \item The edges alternate between $\Diff_t$ and $\Diff_c$, i.e.,
    \[(v_i, v_{i+1}) \in \Diff_t \;\Rightarrow\; (v_{i+1}, v_{i+2}) \in \Diff_c,\]
    and
    \[(v_i, v_{i+1}) \in \Diff_c \;\Rightarrow\; (v_{i+1}, v_{i+2}) \in \Diff_t,\]
    for all $i \in \{1,\dots,k-1\}$.
        \item If $v_1 \neq v_{k+1}$, then both $v_1$ and \(v_{k+1}\) have degree 1 in $\Diff$, and the sequence forms an open path. If $v_1 = v_{k+1}$, the sequence forms a cycle.
    \end{itemize}
\end{definition}


\addtocounter{example}{-1}
\begin{example}[Continued]
    In Figure~\ref{fig:matching-example} (third panel), the disagreement decomposes into two alternating cycles, $\{1,6,2,7,1\}$ and $\{3,8,4,9,3\}$. One can also visualize alternation by noting that adjacent edges in each cycle are with different colors, which correspond to different sets , $\Diff_t$ and $\Diff_c$. In this instance, the decomposition into these two cycles seems to be unique. 
\end{example}

The endpoint condition in Definition~\ref{def:alternating-path} ensures that alternating paths cannot be extended further within $\Diff$. Without this condition, any connected subsequence of vertices along a maximal alternating path or cycle would also qualify as an alternating path, which would prevent a unique decomposition. With the condition in place, we obtain the following uniqueness result.


\begin{proposition}[Uniqueness of Decomposition] \label{prop:unique-decomposition} Let $\sM^t$ and $\sM^c$ be two one-to-one matchings. Then the disagreement set $\Diff$ 
admits a \textbf{unique} decomposition into a disjoint collection of alternating paths and alternating cycles. Furthermore, every connected component of the graph induced by $\Diff$ is either an alternating path or an alternating cycle.
\end{proposition}
The proposition follows from two simple observations. First, the connected components of any edge set are uniquely determined. Second, since $\sM^t$ and $\sM^c$ are one-to-one matchings, each agent is incident to at most one edge in $\Diff_t$ and at most one edge in $\Diff_c$. Consequently, every agent in any connected component of the graph induced by $\Diff$ has maximum degree two, and its edges necessarily alternate between $\Diff_t$ and $\Diff_c$, forming either an alternating path or an alternating cycle. We let $\Path:=\{\path_1,\cdots, \path_m\}$ denote the unique collection of alternating paths and cycles obtained from decomposing $\Diff$. With mild abuse of notation, we may write $(i,j)\in\Path$ to indicate that $i$ and $j$ are adjacent vertices within the one component pf $\Path$. Additionally, a useful property of alternating cycles we want to highlight is that they always have even length.

\begin{proposition}[Even-Length Cycles]\label{prop: cycle-length}
    Let $\sM^t$ and $\sM^c$ be two one-to-one matchings. For any $\path_i=\{v_{i,1},\cdots,v_{i,k(i)+1}\}\in \Path$ that is a cycle, $k(i)$ must be an even integer with $k(i)\ge 4$.
\end{proposition}
If $k(i)$ were odd, then by the definition of alternation, the first and last edges $(v_{i,1}, v_{i,2})$ and $(v_{i,k(i)}, v_{i,k(i)+1})$ would both belong to $\Diff_t$ or both belong to $\Diff_c$. Since $v_{i,1} = v_{i,k(i)+1}$ in a cycle, this would imply that $v_{i,1}$ is incident to two distinct edges from $\Diff_t$ (or from $\Diff_c$), contradicting the one-to-one matching property that each agent can be matched with at most one partner in $\sM^t$ and in $\sM^c$. Hence, $k(i)$ must be even. The case $k(i)=2$ is impossible because it would require two distinct agents simultaneously matched to each other in both matchings, contradicting the definition of a disagreement edge. Therefore, $k(i)\ge 4$.

We first note that matching interference is fully captured within each alternating component. Adjacent edges on an alternating path cannot be realized simultaneously, and no such interference arises between different components in $\Path$. Hence, randomization can be performed independently across cycles and paths, and we describe AP design on a generic component $\path_i=\{v_{i,1},\cdots,v_{i,k(i)+1}\} \in \Path$. Foe each edge $(v_{i,j},v_{i,j+1})$, we define a binary indicator $W_{i,j}\in\{0,1\}$, where $W_{i,j}=1$ indicates that the match is realized in the experiment. The randomization proceeds sequentially along the path/cycle. More specifically, the randomization of $W_{i,j}$ is conditional on $W_{i,j-1}$. If $W_{i,j-1}=1$, $W_{i,j}$ has to be zero, reflecting that adjacent matches cannot both be realized due to capacity constraints. If $W_{i,j-1}=0$, $W_{i,j}$ can be randomized as 1 with probability $p$. More formally, we define our AP design in the following definition.

\begin{definition}[Alternating Path Randomized Design (AP Design)]
\label{def:APRD}
Let $\Path = \{\path_1,\cdots,\path_i,\cdots,\path_m\}$ be the alternating path/cycle decomposition of $\Diff$. AP design generates a binary selection vector $\mW = \{W_{i,j}\}_{i\in [m],j\in[k(i)]}$ with the following rules:

\textbf{(i) Independence across paths and cycles.} Randomization is performed independently across paths/cycles $\path_i$.

\textbf{(ii) Path case.} If $\path_i = (v_{i,1},\dots,v_{i,k(i)+1})$ is an open path, then
\begin{equation}\label{eq:first-edge}
    \prob(W_{i,1}=1)=\frac{p}{1+p}.
\end{equation}
and for $j=2,\dots,k(i)$,
\begin{equation}\label{eq:conditional-randomization}
\mathbb{P}(W_{i,j}=1 \mid W_{i,j-1}=0) = p,
\qquad
\mathbb{P}(W_{i,j}=1 \mid W_{i,j-1}=1) = 0.    
\end{equation}

\textbf{(iii) Cycle case.} If $\path_i = (v_{i,1},\dots,v_{i,k(i)+1})$ is a cycle, for $1\le j\le k(i)-1$, $W_{i,j}$ follows the same rules in \Eqref{eq:first-edge} and \Eqref{eq:conditional-randomization}. The final edge (i.e., $j=k(i)$) is set deterministically:
\[
\mathbb{P}(W_{i,k(i)} = 1 \mid W_{i,1}=0, \; W_{i,k(i)-1}=0) = 1,
\]
and $W_{i,k(i)}=0$ otherwise.

\end{definition}

There are several remarks regarding this randomization procedure. First, although $W_{i,k(i)}$ is set deterministically conditional on $W_{i,1}$ and $W_{i,k(i)-1}$ in the cycle case, it remains a random variable because $W_{i,1}$ and $W_{i,k(i)-1}$ are themselves random. Second, we do not introduce an additional randomization step for $W_{i,k(i)}$ because this final decision does not affect the feasibility of any other edges, and setting it deterministically allows us to include as many matches as possible. Similar conditional randomization ideas have appeared, for example, in \citet{ni2025enhancing}, though in a different context for improving switchback experiments. For notational convenience, we use $\mW_i$ to denote the vector of indicators on component $\path_i$ and $\mW$ for the full assignment on the disagreement set $\Diff$. The realized selection $\mW$ induces the experimental matching $\sM^{\text{AP}}$. Match pairs outside $\Diff$ can be implemented arbitrarily, as they do not affect the estimation of the average treatment effect; however, implementing them can be useful if one is interested in estimating $\bar{Y}_t$ or $\bar{Y}_c$ themselves.


For the estimation, given any realization of the assignment $\vw$, we use the Horvitz-Thompson estimator for the average treatment effect:
\begin{equation}\label{eq:HT-estimator}
    \AP=\frac{1}{N}\sum_{\path_i \in \Path} \left( \sum_{\substack{ (v_{i,j},v_{i,j+1}) \in \\  \path_i \cap \Diff_t}}\frac{\sI\{w_{i,j}=1\}\cdot Y_{v_{i,j},v_{i,j+1}}}{\prob(W_{i,j}=1)}-\sum_{ \substack{ (v_{i,j},v_{i,j+1}) \in \\  \path_i \cap \Diff_c}}\frac{\sI\{w_{i,j}=1\}\cdot Y_{v_{i,j},v_{i,j+1}}}{\prob(W_{i,j}=1)}\right).
\end{equation}
Note that if $\sI\{w_{i,j}=1\}=1$, $Y_{v_{i,j},v_{i,j+1}}$ is observed and if $\sI\{w_{i,j}=1\}=0$, though $Y_{v_{i,j},v_{i,j+1}}$ is unobservable,  $\sI\{w_{i,j}=1\} \cdot Y_{v_{i,j},v_{i,j+1}}$ equals to zero anyway. Therefore, $\AP$ is well-defined under the AP design. For notational convenience, we write
\begin{equation*}
    \AP= \frac{1}{N}\sum_{\path_i \in \Path}\EPTE_i\text{ and } \PTE_i=\sum_{ \substack{ (v_{i,j},v_{i,j+1}) \in \\  \path_i \cap \Diff_t}} Y_{v_{i,j},v_{i,j+1}}-\sum_{ \substack{ (v_{i,j},v_{i,j+1}) \in \\  \path_i \cap \Diff_c}} Y_{v_{i,j},v_{i,j+1}},
\end{equation*}
where $\PTE_i$ represents the corresponding fixed non-random path-level treatment effect and $\EPTE_i$ is its estimation. In the following lemma, we compute the non-conditioned probability of each $W_{i,j}=1$, $\prob(W_{i,j}=1)$ for each edge undet the AP design.


\begin{lemma} \label{lemma:uncondition-prob}
    For any $\path_i=(v_{i,1},\cdots,v_{i,k(i)+1})\in\Path$, the unconditional probability of $W_{i,j}=1$ under AP design satisfies,
    \begin{equation}\label{eq:uncondition-prob}
        \prob(W_{i,j}=1) =
\begin{cases}
\frac{1-(-p)^{j-1}}{(1+p)^2}, & \text{if } \path_i \text{ is a cycle and } j=k(i), \\
\frac{p}{p+1}, & \text{otherwise}.
\end{cases}
    \end{equation}
\end{lemma}
The lemma shows that only the final edge of an alternating cycle has a different unconditional selection probability. This reflects the fact that, in a cycle, feasibility constraints require the last edge to depend on both its predecessor and the first edge. Moreover, the choice of the initial probability $\mathbb{P}(W_{i,1}=1)={p}/{(1+p)}$ in \Eqref{eq:first-edge} yields the clean closed-form expression in \Eqref{eq:uncondition-prob}.


Since $\prob(W_{i,j}=1)$ is strictly positive whenever $p>0$, every edge in $\Diff$ has positive selection probability. Under the AP design, the Horvitz-Thompson estimator is therefore unbiased in the standard design-based sense.

\begin{proposition}[Unbiasedness of the Horvitz-Thompson estimator] For any $p\in(0,1)$, the Horvitz-Thompson estimator $\AP$ is unbiased for ATE, i.e., $\E[\AP]=\ATE$. \label{prop:unbiasedness}
\end{proposition}

\section{Variance Analysis and Optimal Design}
We now analyze the variance of the Horvitz--Thompson estimator under the AP design and study how to choose the design parameter $p$ (i.e., the conditional probability)  to improve efficiency. The first key observation is that randomization is independent across alternating paths and cycles, which yields a path-wise variance decomposition.

\begin{proposition}[Variance Decomposition]\label{prop:variance-decompostition}
    Under the AP design, the variance of $\AP$ can be decomposed into the sum of the variance of the path-wise estimator $\EPTE_i$, i.e., 
    \begin{equation*}
        \Var (\AP)=\frac{1}{N^2} \sum_{\path_i\in \Path} \Var (\EPTE_i).
    \end{equation*}
\end{proposition}
With Proposition \ref{prop:variance-decompostition}, we can zoom in to the variance of one single alternating path/cycle.

\begin{theorem} \label{thm:variance-decompose}
For $\path_i=\{v_{i,1},\cdots,v_{i,k(i)+1}\}\in\Path$, if $\path_i$ is a path rather than a cycle (i.e., $v_{i,k(i)+1}\not= v_{i,1}$), we have
\begin{equation}\label{eq:variance-of-path}
    \Var(\EPTE_i)= \frac{1}{p} \sum_{j=1}^{k(i)} Y_{v_{i,j},v_{i,j+1}}^2 + 2 \sum_{1\le j<q \le k(i)} p^{q-j-1} Y_{v_{i,j},v_{i,j+1}}Y_{v_{i,q},v_{i,q+1}}.
\end{equation}
If $\path_i$ is a cycle rather than a path, we have
  \begin{align}
    &\Var(\EPTE_i) = \frac{1}{p} \sum_{j=1}^{k(i)-1}  Y_{v_{i,j},v_{i,j+1}}^2 + 2 \sum_{1\le j<q \le k(i)-1} p^{q-j-1} Y_{v_{i,j},v_{i,j+1}}Y_{v_{i,q},v_{i,q+1}}+ \frac{p^2+2p-p^{k(i)-1}}{1+p^{k(i)-1}} Y_{v_{i,k(i)},v_{i,k(i)+1}}^2\notag \\
    &+ 2 \sum_{1\le j \le k(i)-1} (-1)^j \left(\frac{(1-(-p)^{k(i)-1-j})(1-(-p)^{j-1})}{(1+p^{k(i)-1})}-1\right)Y_{v_{i,j},v_{i,j+1}}Y_{v_{i,k(i)},v_{i,k(i)+1}}.\label{eq:variance-of-cycle}
\end{align}  
\end{theorem}

The cycle expression differs from \Eqref{eq:variance-of-path} only through the contribution of the final edge. Along an alternating path, correlations between edge selections decay geometrically with distance. In contrast, in the cycle case the last edge is also constrained by the first edge, closing the loop and inducing additional dependence between the final edge and all preceding edges. This additional dependence gives rise to the extra terms in \Eqref{eq:variance-of-cycle}.



Note that the variance expressions depend on the potential outcomes $Y_{i,j}$, for which we have not imposed any modeling assumptions. To study the order of magnitude of the variance and to facilitate comparison across designs, we introduce a boundedness assumption on potential outcomes below.


\begin{assumption}[Bounded Potential Outcomes]\label{assumption:bounded}
There exists a constant $B>0$ such that for all $(i,j)\in\sM$, the potential outcome satisfies $Y_{i,j}\in[0,B]$.
\end{assumption}
Assumption~\ref{assumption:bounded} is mild in practice, as the upper bound $B$ may be large and need not be known to the experimenter. Similar boundedness conditions are standard in the design-based causal inference literature (see, e.g., \citealt{bojinov2023design, chattopadhyay2023design, chen2025efficient}). Under this assumption, we can study the worst-case variance of the AP design, defined as the largest variance attainable over all potential outcomes satisfying the bound. Formally, for a fixed design parameter $p$ and alternating path/cycle decomposition $\Path$, we consider
\begin{equation}\label{eq:max-variance}
    \max_{\substack{ Y_{i,j}\in [0,B]  \\ \forall (i,j)\in\sM }}\Var (\AP) = \frac{1}{N^2} \max_{\substack{ Y_{i,j}\in [0,B]  \\ \forall (i,j)\in\sM }} \sum_{\path_s\in \Path} \Var (\EPTE_s) = \frac{1}{N^2}  \sum_{\path_s\in \Path} \max_{\substack{ Y_{i,j}\in [0,B]  \\ \forall (i,j)\in\path_s }} \Var (\EPTE_s),
\end{equation}
where the second equality follows from Proposition~\ref{prop:variance-decompostition}, since randomization is independent across components. Fortunately, The maximization problem in \Eqref{eq:max-variance} admits a simple solution that the worst-case variance is achieved when all relevant potential outcomes attain their upper bound $B$.
\begin{proposition}\label{prop:max-variance}
    The solution to the maximization problem of \Eqref{eq:max-variance} is $Y_{i,j}=B$ for all $(i,j)\in \sM$. Particularly, under Assumption \ref{assumption:bounded}, all the correlation terms in \Eqref{eq:variance-of-path} and \Eqref{eq:variance-of-cycle} are positive.
\end{proposition}
Proposition~\ref{prop:max-variance} implies that all coefficients multiplying the potential outcomes in Theorem~\ref{thm:variance-decompose} are nonnegative, including those associated with cross-edge correlations. Consequently, the largest variance is achieved when all potential outcomes attain their maximal value $B$. The intuition is as follows. Along an alternating component $\path_i=\{v_{i,1},\cdots,v_{i,k(i)+1}\}$, the adjunct two indicators $W_{i,j}$ and $W_{i,j+1}$ are negatively dependent due to feasibility constraints that if one edge is realized, the adjacent edge must be skipped. However, adjacent edges belong to different matchings (treatment v.s. control), so their contributions enter the estimator with opposite signs. These two effects combine to produce positive covariance terms in the variance expression. Related phenomena have been observed in switchback experiments with temporal structure (e.g., \citealt{bojinov2023design, chen2025efficient}), although the underlying dependence in our setting arises from matching interference rather than time. 
Building on Proposition~\ref{prop:max-variance}, we now characterize the worst-case variance contribution of a single alternating path or cycle under the AP design.


\begin{theorem}\label{thm:variance-upper-bound}
    Suppose Assumption~\ref{assumption:bounded} holds, for $\path_s=\{v_{s,1},\cdots,v_{s,k(s)+1}\}\in\Path$, if $\path_s$ is a path, we have
    \begin{equation} \label{eq:path-variance-upper-bound}
    \max_{\substack{ Y_{i,j} \in [0,B]  \\ \forall (i,j)\in\path_s }} \Var (\EPTE_s)= \left(\frac{1}{p}+\frac{2}{1-p}\right) B^2 k(s) + \frac{2B^2(p^{k(s)}-1)}{(1-p)^2}.
\end{equation}
If $\path_s$ is a cycle, we have
\begin{align}
     \max_{\substack{ Y_{i,j} \in [0,B]  \\ \forall (i,j)\in\path_s }} \Var (\EPTE_s)&= \left(\frac{1}{p}+\frac{2}{1-p}\right) B^2 (k(s)-1) +\frac{2B^2(p^{k(s)-1}-1)}{(1-p)^2}\notag\\
     &\qquad\qquad+\frac{B^2(4+2p-p^2-p^3-p^{k(s)-2}(2+p+p^2))}{(1-p)(1+p^{k(s)-1})}.\label{eq:cycle-variance-upper-bound}
\end{align}
Moreover, for both altering paths and cycles, we can have
\begin{equation*}
    \lim_{k(s)\rightarrow \infty} \frac{\max_{\substack{ Y_{i,j}\le [0,B]  \\ \forall (i,j)\in\path_s }} \Var (\EPTE_s)}{k(s)} = \frac{B^2(1+p)}{p(1-p)}.
\end{equation*}
Equivalently, under the worst case, $\Var (\EPTE_s)=\frac{B^2(1+p)}{p(1-p)} k(s) +o(k(s))$.
\end{theorem}

Theorem~\ref{thm:variance-upper-bound} shows that for each path/cycle, the variance of the estimator grows only linearly in its length. Therefore, aggregating over components yields
\begin{equation*}
    \Var (\AP) = \frac{1}{N^2}\left(\sum_{\path_s\in \Path} \frac{B^2(1+p)}{p(1-p)} k(s) +o(k(s)) \right) = \frac{B^2(1+p)}{p(1-p)}\cdot \frac{1}{N}+o\left(\frac{1}{N}\right),
\end{equation*}
where we have used $\sum_{s}k(s)=N$ for one-to-one matching. 
Thus, the variance of the AP estimator decreases at rate $1/N$, despite the dependence induced by matching interference. For comparison, under the naive design in Proposition~\ref{prop:naive-design}, the worst-case variance $\Var(\Naive)$ is $4B^2$, which does not vanish as $N\rightarrow\infty$. Finally, Theorem~\ref{thm:variance-upper-bound} implies that although the indicators $W_{i,j}$ are dependent along each component, the induced correlations decay sufficiently fast with distance that asymptotic efficiency is preserved.


\subsection{Optimal AP Design under Minimax Rule}

With Theorem~\ref{thm:variance-decompose} in hand, we now turn to optimizing the randomization probability $p$. In the previous sections, for the ease of presentation, we set all the paths and cycles to share a common value of $p$. However, since randomization across alternating paths is independent, each alternating path/cycle $\path_s$ can in principle adopt its own design parameter $p_s$. In this section, we therefore concentrate on optimizing the probability for one path/cycle. Following a common practice in decision theory (\citealt{berger2013statistical}) and experimental design (\citealt{bojinov2023design}), we formulate the design problem into a minimax problem, under which we seek to minimize the worst case risk, i.e.,
\begin{equation}\label{eq:minimax}
   \min_{p}  \max_{\substack{ Y_{i,j}\in [0,B]  \\ \forall (i,j)\in\path_s }} \E\left[(\EPTE_s-\PTE)^2\right]=  \min_{p}  \max_{\substack{ Y_{i,j}\in [0,B]  \\ \forall (i,j)\in\path_s }} \Var (\EPTE_s).
\end{equation}
The equation holds since our $\EPTE_s$ is unbiased. Thus, we are actually trying to minimize the RHS of \Eqref{eq:path-variance-upper-bound} or \Eqref{eq:cycle-variance-upper-bound} with respect to $p$. For a path/cycle of finite length $k(s)$, a closed-form minimizer of  \Eqref{eq:path-variance-upper-bound} or \Eqref{eq:cycle-variance-upper-bound} is generally not available due to the mixed polynomial–exponential structure of the variance expression. However, when the path is long, the normalized worst-case variance is asymptotically linear in $k(s)$ with leading coefficient $(1+p)/(p(1-p))$. This yields a limiting optimal probability.
\begin{corollary}
    Let $p^*_{k(s)}$ denote the optimal solution $p$ to \Eqref{eq:minimax} of a path with length $k(s)$. Then, we have that $\lim_{k(s)\rightarrow\infty}p^*_{k(s)}=\sqrt{2}-1\approx 0.4142$.
\end{corollary}
Thus, for long alternating paths, the AP design asymptotically favors a conditional randomization probability strictly below $1/2$. Intuitively, choosing $p<1/2$ reflects the fact that we are not only minimizing the variance contribution of the current edge, but must also account for downstream edges and control the propagation of interference along the path. 

Table~\ref{tab:p_star}reports numerical solutions for the optimal $p^*_{k(s)}$ under both the path case and the cycle case for various values $k(s)$. When $k(s)=1$, there is no interference since only one match exists, so we can set $p=1$, and the match is always included. For this reason, the table begins at $k(s)=2$. In the cycle case, as we have discussed in Proposition~\ref{prop: cycle-length}, the minimum feasible length of any cycle is $k(s)=4$ and $k(s)$ must be even. For both paths and cycles, the optimal value satisfies $p_{k(s)}^*=1$ whenever $k(s)\le 4$. In these short cases, the minimax-optimal strategy coincides with applying the naive design on the component that with probability $1/2$ one adopts all treatment edges and with probability $1/2$ one adopts all control edges. As the length increases, the AP design gradually yields lower worst-case variance compared to the naive design. The case $k(s)=4$ therefore serves as the critical breakpoint beyond which a carefully chosen $p<1$ is beneficial. Consistent with the asymptotic analysis, when $k(s)\ge 50$ the optimal probability $p_{k(s)}^*$ has effectively stabilized near $\sqrt{2}-1$. 

\begin{table}[h!]
\centering
\caption{Numerical minimizers $p^*_{k(s)}$ for Path and Cycle Cases.}
\label{tab:p_star}
\begin{tabular}{c c c c c}
\toprule
 & \multicolumn{2}{c}{\textbf{Path Case}} 
 & \multicolumn{2}{c}{\textbf{Cycle Case}} \\ 
\cmidrule(lr){2-3} \cmidrule(lr){4-5}
 $k(s)$ & $p^*_{k(s)}$ 
 & $ \frac{\max Var(\hat{\Gamma}_s)}{k(s)}\;(B=1)$
 & $p^*_{k(s)}$ 
 & $ \frac{\max Var(\hat{\Gamma}_s)}{k(s)}\;(B=1)$ \\ 
\midrule
2   & $1$    & $2$       & ---         & --- \\ 
4   & $1$    & $4$     & $1$     & $ 4$ \\
5   & 0.61552 & 4.35964 & --- & --- \\
6   & 0.55263    & 4.66204       & 0.46505     & 5.13646 \\
10  & 0.47272    & 5.18951       & 0.42952     & 5.44095 \\
50  & 0.42296    & 5.71012       & 0.41658     & 5.75173 \\
100 & 0.41847    & 5.76972      & 0.41537     & 5.79011 \\
1000& 0.41463    & 5.82259       & 0.41433     & 5.82460 \\
$\vdots$ & $\vdots$ & $\vdots$   & $\vdots$    & $\vdots$ \\
$\infty$ & $\sqrt{2}-1$ & $3+2\sqrt{2}$ & $\sqrt{2}-1$ & $3+2\sqrt{2}$ \\
\bottomrule
\end{tabular}
\end{table}

\section{Asymptotic Inference}
\begin{figure}[th]
    \centering
    \includegraphics[width=0.8\linewidth]{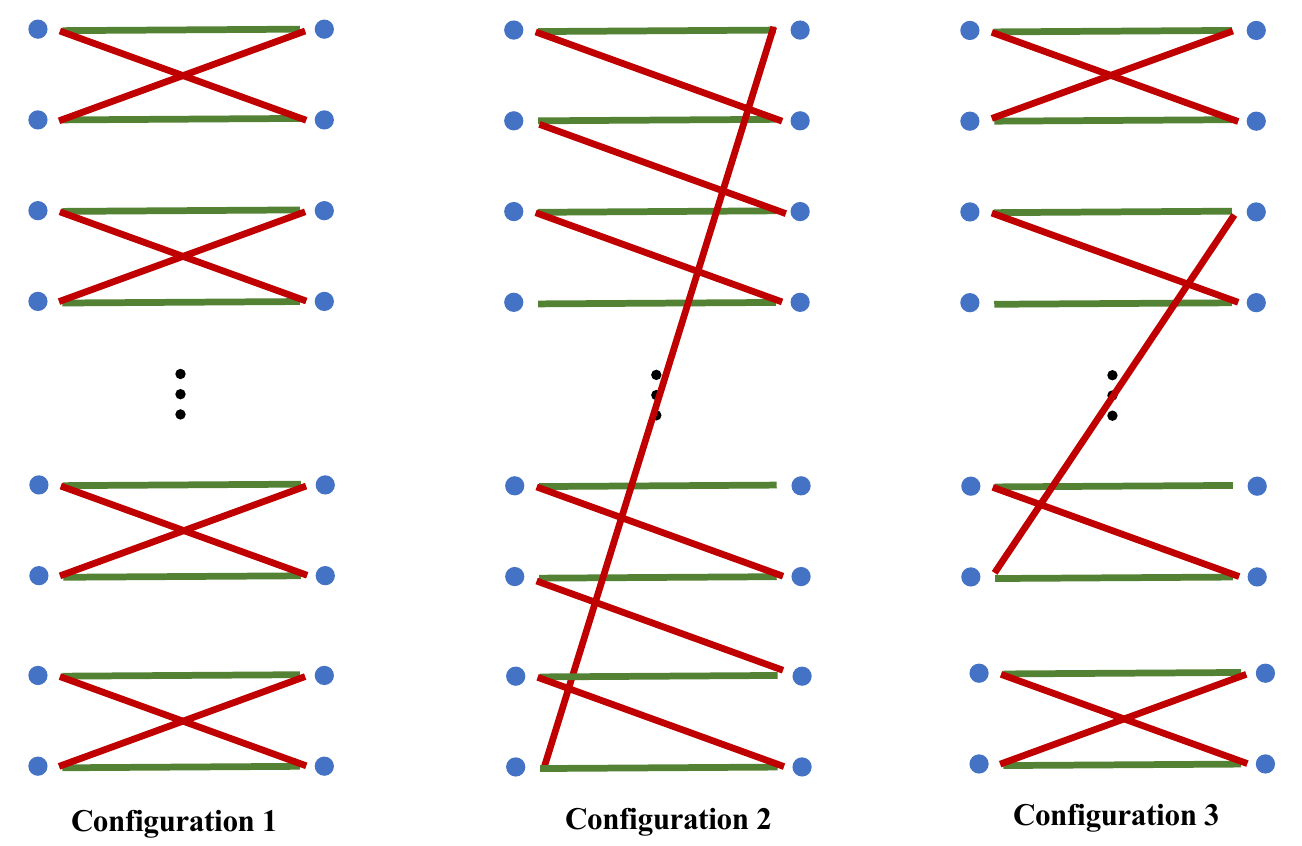}
    \caption{Three Possible Configurations of Disagreement Sets}
    \label{fig:config-disagreement-set}
\end{figure}

In this section, we present the asymptotic inference result of our estimator $\AP$. 
\begin{theorem}[Central Limit Theorem]\label{thm:clt}
    For AP design, if Assumption \ref{assumption:bounded} holds, the size of the disagreement set goes to infinity i.e., $|\Diff|\rightarrow\infty$ and $N^2\Var(\AP)\rightarrow \infty$, then, as $N\rightarrow\infty$,
    \begin{equation*}
        \frac{\AP-\ATE}{\sqrt{\Var(\AP)}} \xrightarrow{d} \mathcal{N}(0,1),
    \end{equation*}
    where $\xrightarrow{d}$ denotes convergence in distribution.
\end{theorem}
Although Theorem~\ref{thm:clt} resembles a standard central limit result, it does not appear to follow from a direct application of well-known CLTs. The difficulty arises from the potentially complex structure of the alternating path/cycle decomposition $\Path$. Figure~\ref{fig:config-disagreement-set} illustrates three qualitatively distinct configurations that may occur. Configuration 1 corresponds to the case where all components in $\Path$ are uniformly small. In this regime, the classical Lindeberg–Feller CLT (e.g., \citealt{feller1991introduction}) applies directly, as dependence is localized and vanishes in aggregate. Configuration 2 corresponds to the opposite extreme, where there exists a single (or a small fixed number of) components of diverging length. In this regime, we cannot appeal to i.i.d. or Lindeberg–Feller arguments, since we impose no distributional model on $Y_{i,j}$. However, on an infinitely long alternating path, the sequence $W_{i,j}$ behaves (up to boundary effects) as a Markov chain with transition matrix:
\[
\mP = \begin{pmatrix}
1 - p & p \\
1 & 0
\end{pmatrix},
\]
so that nonstationary $\alpha$-mixing CLT results (e.g., \citealt{bradley2017central}) may be invoked.

The substantial challenge arises in Configuration 3, where $\Path$ contains a mixture of finitely long and infinitely long components. In this case, no single CLT applies uniformly. Moreover, as $N$ grows, the configuration itself may fluctuate among the three types, and even within Configuration 3 the relative proportions of short and long components need not converge. At a high level, our proof partitions $\Path$ into two groups: a long-path group $\Path_L$ and a short-path group $\Path_S$. We define the path/cycle whose variance $\Var(\EPTE_i)$ is larger than $(\sum_i \Var(\EPTE_i))^{2/3}$ as the long path, otherwise as a short path.  Both $|\Path_L|$ and $|\Path_S|$ may diverge, and neither group is assumed to be stabilized as $N$ grows. sing the Bolzano–Weierstrass theorem together with the subsequence principle (Theorem 2.6 in \citealt{billingsley2013convergence}), we show that every subsequence of contributions from $\Path_L$ and $\Path_S$ admits a further subsequence along which a normal limit holds. On a side note, for infinite components, the only deviation from the Markovian structure comes from the final edge of a cycle, whose boundary effect vanishes relative to the component length. For finite components, the interior dependence does not affect the limit as long as the component size is uniformly bounded. Together, these arguments establish the asymptotic normality of $\AP$ under the stated conditions.

A practical caveat here is that the closed-form expressions in \Eqref{eq:variance-of-path} and \Eqref{eq:variance-of-cycle} are not identifiable from observed data. In particular, terms of the form $Y_{v_{i,j},v_{i,j+1}}Y_{v_{i,q},v_{i,q+1}}$ with $q=j+1$ cannot be observed, since matching interference guarantees that two consecutive edges along an alternating component are never simultaneously realized. As a consequence, the true variance cannot be estimated directly. To construct confidence intervals and yield implementable inference, we therefore derive an estimable upper bound on the true variance. Using the Cauchy–Schwarz inequality, we obtain an upper bound for the path case that depends only on observable quantities and can be estimated unbiasedly from the realized data. The cycle case follows the same principle, and the explicit expression is deferred to Appendix~\ref{app:sec:var-upper-bound}.
\begin{proposition}\label{prop:estimated-variance}
    For $\path_i=\{v_{i,1},\cdots,v_{i,k(i)+1}\}\in\Path$, if $\path_i$ is a path, we have the following upper bound $\tilde{\sigma}^2_i$ of $\Var(\EPTE_i)$,
\begin{equation}
    \tilde{\sigma}^2_i := \left(\frac{1}{p}+1\right) \sum_{j=1}^{k(i)} Y_{v_{i,j},v_{i,j+1}}^2+\sum_{j=2}^{k(i)-1} Y_{v_{i,j},v_{i,j+1}}^2 + 2 \sum_{\substack{1\le j \le k(i)-2 \\ j+2\le q \le k(i)}} p^{q-j-1} Y_{v_{i,j},v_{i,j+1}}Y_{v_{i,q},v_{i,q+1}}.
\end{equation}
Moreover, $\tilde{\sigma}^2_i$ can be estimated by
\begin{align*}
    \hat{\sigma}^2_i&:= \left(\frac{1}{p}+1\right) \sum_{j=1}^{k(i)} \frac{\sI\{w_{i,j}=1\}\cdot Y^2_{v_{i,j},v_{i,j+1}}}{\prob(W_{i,j}=1)}+\sum_{j=2}^{k(i)-1} \frac{\sI\{w_{i,j}=1\}\cdot Y^2_{v_{i,j},v_{i,j+1}}}{\prob(W_{i,j}=1)}\\
    &\qquad\qquad\qquad + 2 \sum_{\substack{1\le j \le k(i)-2 \\ j+2\le q \le k(i)}} p^{q-j-1} \frac{Y_{v_{i,j},v_{i,j+1}}Y_{v_{i,q},v_{i,q+1}}\sI\{w_{i,j}=1,w_{i,q}=1\}}{\prob(W_{i,j}=1,W_{i,q}=1)},
\end{align*}
where $\prob(W_{i,j}=1)=p/(p+1)$ and $\prob(W_{i,j}=1,W_{i,q}=1)=(p^2-(-p)^{q-j+1})/(p+1)^2$. The estimator $\hat{\sigma}^2_i$ is an unbiased estimator of $\tilde{\sigma}^2_i$, i.e., $\E[\hat{\sigma}^2_i]=\tilde{\sigma}^2_i$.
\end{proposition}
There are alternative ways to obtain conservative variance bounds. For example, we can bound all the cross term $Y_{v_{i,j},v_{i,j+1}}Y_{v_{i,q},v_{i,q+1}}$ by $1/2(Y_{v_{i,j},v_{i,j+1}}^2+Y_{v_{i,q},v_{i,q+1}}^2)$, regardless of whether they are observable or not. In addition, we note that Proposition \ref{prop:estimated-variance} does not rely on the boundedness assumption (Assumption~\ref{assumption:bounded}). If we do wish to exploit boundedness, the unobserved outcomes can always be bounded by the constant $B$. Using Proposition \ref{prop:estimated-variance},  $\sqrt{\Var(\AP)}$ can be estimated slightly conservatively via $\hat{\sigma}^2:={1}/{N^2} \sum_{\path_i\in \Path} \hat{\sigma}^2_i$. Then, we can form an approximate asymptotic $\alpha$-level confidence interval for the AP estimator by
\[\left[\AP-c_{\frac{1+\alpha}{2}}\hat{\sigma},\AP+c_{\frac{1+\alpha}{2}}\hat{\sigma}\right],\]
where $c_{(1+\alpha)/2}$ is the $(1+\alpha)/2$ quantile of the standard normal distribution.

\section{Many-to-One Matching}
We now extend the AP design to two-sided many-to-one environments, in which agents on one side of the market may match with multiple agents on the other side. Such settings arise frequently in applications, including student–school assignments (\citealt{chen2016school}), split-liver transplantation (\citealt{tang2025multi, tang2025split}), ridesharing and carpooling platforms (\citealt{castillo2025matching}), and refugee resettlement (\citealt{bansak2024dynamic}). In each of these examples, suppliers possess a capacity that constrains the number of demands they may serve, while each demand may be served by at most one supplier.

Formally, let the set of suppliers be $\sS:=\{s_1,\cdots,s_N\}$ and the set of demands be $\sD:=\{d_1,\cdots,d_M\}$. Each supplier $s_i\in\sS$ has capacity $C_i$, meaning it may be matched with at most $C_i$ demands, while each demand may be matched with at most one supplier. For simplicity of exposition we assume homogeneous capacities $C_i=C_0$ for all $s_i$, though our results extend to heterogeneous capacities without conceptual difficulty. Let $m_{i,j}\in\{0,1\}$ indicate whether supplier $i$ is matched with demand $j$. A feasible matching $\sM$ in many-to-one matching with capacity constraint is then defined by
\begin{equation}\notag
    \sM=\left\{(i,j): i\in [N]; j\in [M]; m_{i,j}=1; \sum_{j\in [M]}m_{i,j}\le C_0, \;\forall i\in[N] ; \sum_{i\in[N]}m_{i,j}\le 1, \;\forall j\in [M] \right\},
\end{equation}
and we let $\gM$ denote the set of all feasible many-to-one matchings.

Our inferential goal remains the same: given two feasible matchings $\sM^t \in \gM$ and $\sM^c \in \gM$, we seek to experimentally compare their average performance without imposing outcome or behavioral models. For a potential outcome $Y_{i,j}$ defined for each feasible pair $(i,j)$, we define the ATE analogously to the one-to-one case as
\begin{equation}\notag
    \ATE=\frac{\sum_{(i,j)\in \sM^t} Y_{i,j}}{C_0 N}-\frac{\sum_{(i,j)\in \sM^c} Y_{i,j}}{C_0 N}.
\end{equation}
The normalization by $C_0 N$ reflects the maximum number of possible matches. Normalizing by the number of demand $M$ is also reasonable and would yield an equivalent estimand up to scale. Finally, following the philosophy of \eqref{eq:diff-1-1}, we define the disagreement set as
\begin{equation}\notag
    \Diff =  \{(i,j)|(i,j)\in\sM^t;(i,j)\notin\sM^c \} \cup \{(i,j)|(i,j)\in\sM^c;(i,j)\notin\sM^t \} .
\end{equation}

If we can identify a collection of alternating paths and cycles whose randomized realizations form a feasible matching in $\gM$, then the AP design and the associated inference results from the one-to-one setting continue to apply without modification. Thus, the new challenges introduced by the many-to-one setting arise not from inference, but from ensuring that the decomposition of the disagreement set admits a feasible randomized implementation that respects all capacity constraints. In the one-to-one case this issue does not arise. Proposition~\ref{prop:unique-decomposition} guarantees that the alternating path/cycle decomposition is unique, and every alternating component is automatically feasible under randomization. In contrast, in the many-to-one case, (i) the disagreement set may admit multiple distinct alternating decompositions, and (ii) not all decompositions are feasible, since randomization along certain paths may violate supplier capacities.

We therefore begin by introducing a set of sufficient conditions under which an alternating decomposition of the disagreement set yields a feasible randomized implementation under the AP design for the many-to-one matching.

\begin{figure}
    \centering
    \includegraphics[width=0.9\linewidth]{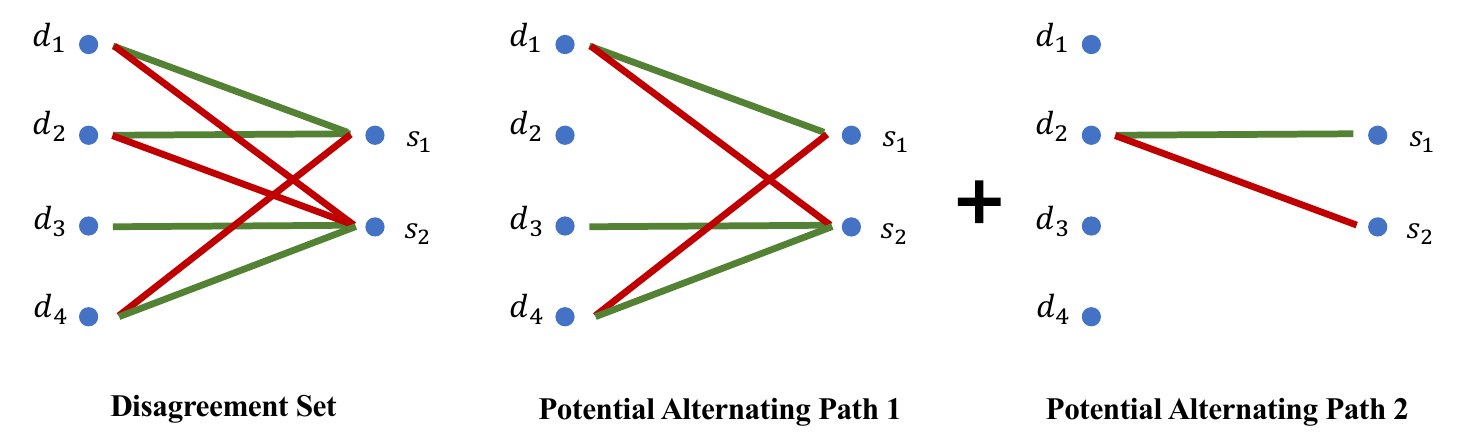}
    \caption{A Potential Decomposition of Alternating Paths and Cycles Without Condition 4.}
    \label{fig:condition-4-counter}
\end{figure}

\begin{theorem}\label{thm:colloction-path}
Let $\Path:=\{\path_1,\cdots, \path_p\}$ be a collection of alternating paths and cycles. If the following conditions hold, then the AP design yields a feasible randomized matching in $\sM$ and the central limit theorem in Theorem~\ref{thm:clt} continues to hold:
    \begin{enumerate}
        \item Each match in $\Diff$ is covered exactly once by $\Path$;
        \item Each supplier appears in at most $C_0$ alternating paths and cycles in $\Path$;
        \item Each demand appears in at most one alternating path or cycle in $\Path$;
        \item Within each $\path_i\in\Path$, no supplier or demand is traversed or visited more than once.
    \end{enumerate}
\end{theorem}
Note that one implicit condition is that each path or cycle must be an alternating path or cycle. The first condition ensures that every disagreement edge is accounted for exactly once. If an edge is omitted or duplicated, the resulting estimator would be biased. Conditions 2–4 jointly guarantee feasibility under the AP design. In particular, Condition 4 implies that any supplier (or demand) appears on a given path at most twice, once as an incoming edge and once as an outgoing edge, and under the AP design at most one of these two edges can be realized due to alternating interference. Combined with Condition 2, this ensures that no supplier exceeds its capacity across all paths. Condition 3 analogously guarantees demand feasibility. The necessity of Condition 4 may not be immediately obvious. Figure~\ref{fig:condition-4-counter} illustrates a decomposition in which $C_0=2$ and Conditions 1–3 are satisfied, but Condition 4 fails because supplier $s_2$ appears three times along the same path. Under the AP design, it is then possible for edges $(d_3,s_2)$ and $(d_4,s_2)$ from Path 1 and $(d_2,s_2)$ from Path 2 to be realized simultaneously, violating the capacity constraint of supplier $s_2$. Condition 4 rules out precisely such cases.


We next describe an efficient procedure for constructing a collection of alternating paths and cycles that satisfies the conditions in Theorem~\ref{thm:colloction-path}. The key idea is to represent the disagreement set as an auxiliary directed graph and reduce the decomposition problem to two classical graph-theoretic tasks: \textit{finding augmenting paths} and \textit{performing Euler–tour decompositions}.



\begin{definition}[Auxiliary Unbalanced Directed Graph]
Given a disagreement set $\Diff$, define its auxiliary directed graph $\gG=(\mathcal{V},\mathcal{E})$ as follows.

\textbf{Vertices.} The vertex set $\mathcal{V}$ consists of:
\begin{enumerate}
    \item Supplier vertices $v_{s_i}$ for each $s_i \in \sS$;
    \item Degree-1 demand vertices $v_{d_j}$ for each $d_j \in \sD$ that is matched with exactly one supplier in $\Diff$.
\end{enumerate}

\textbf{Directed edges.} The edge set $\mathcal{E}$ consists of two types:
\begin{enumerate}
    \item \emph{Supplier-to-supplier edges.} For each demand $d_k$ such that both $(s_i,d_k)$ and $(s_j,d_k)$ lie in $\Diff$, if $(s_i,d_k)\in\sM^t$ and $(s_j,d_k)\in\sM^c$, then include a directed edge from $s_i$ to $s_j$, $(v_{s_i},v_{s_j})$ in $\mathcal{E}$ and label it by $d_k$.
    \item \emph{Supplier–demand edges.} For each degree-1 demand vertex $v_{d_j}$, if $(s_k,d_j)\in\sM^t$ then include a directed edge $(v_{s_k},v_{d_j})$. If $(s_k,d_j)\in\sM^c$ then include $(v_{d_j},v_{s_k})$ instead.
\end{enumerate}
\end{definition}

\begin{figure}[th]
    \centering
    \includegraphics[width=0.9\linewidth]{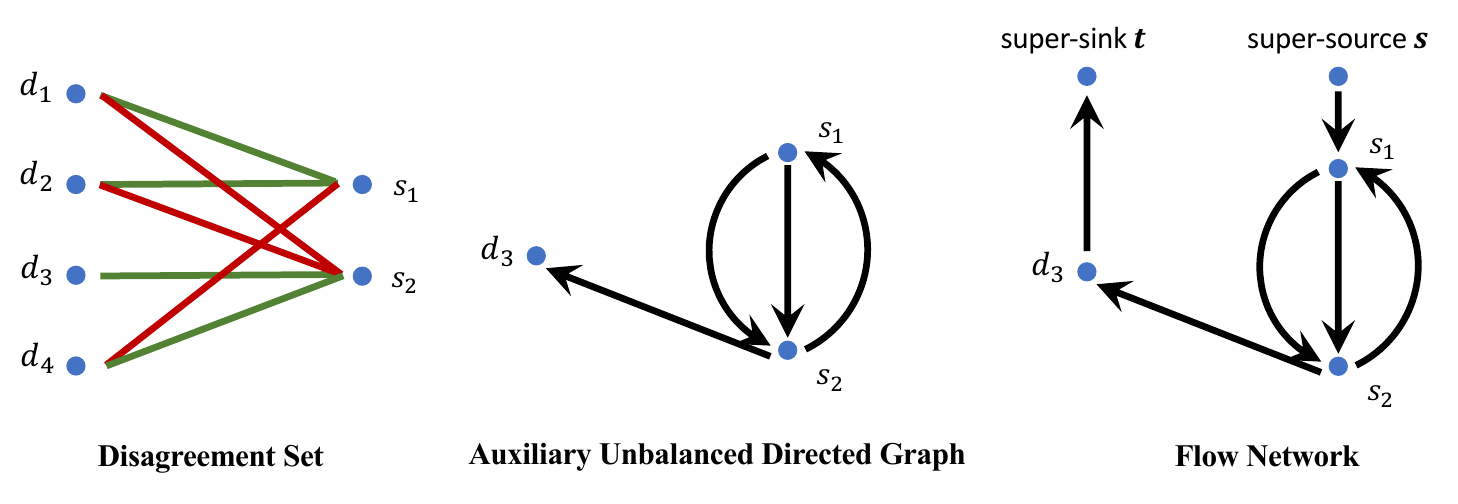}
    \caption{Examples of the auxiliary unbalanced directed graph and the flow graph}
    \label{fig:flow-graph}
\end{figure}

In Figure~\ref{fig:flow-graph}, we showcase the auxiliary unbalanced directed graph of the disagreement set in Figure~\ref{fig:condition-4-counter}. There are several useful observations and properties of the graph that we want to highlight. Each directed edge represents one or two matches in the disagreement set. Demands of degree two in $\Diff$ (i.e., the demands that are matched to two different suppliers in the treatment matching and the control matching) induce supplier-to-supplier edges, while degree-one demands induce supplier–demand edges. Multiple directed edges may exist between the same pair of suppliers, and they may point in the same or opposite directions. For each supplier node in $\gV$,  the number of the outgoing edges (i.e., the out-degree, $\outde(v)$) is equal to the number of matches associated with the supplier in the treatment matching. Correspondingly, the in-degree of each supplier node, $\inde(v)$, is the number of matches in the control matching associated with node. Thus, for each supplier node $v_{s_i}$, neither $\inde(v_{s_i})$ nor $\outde(v_{s_i})$ can be larger than $C_0$. Each degree-1 demand node $v_{d_i}$ has either $\inde(v_{d_i})=1$ or $\outde(v_{d_i})=1$ depending on whether the associated match is in the treatment matching or the control matching. Moreover, the graph is generally unbalanced, meaning that for at least some vertices $v$ such that $\inde(d)\not = \outde(d)$. Every disagreement set $\Diff$ induces a unique auxiliary unbalanced directed graph $\gG$. Conversely, every simple directed path or simple directed cycle in $\gG$ corresponds uniquely to an alternating path or cycle in $\Path$. We denote this inverse mapping by $\gA^{-1}$, so that for any directed path $\pi$ in $\gG$, the object $\gA^{-1}(\pi)$ is an alternating path or cycle in $\Path$.

With the auxiliary graph in place, our task reduces to finding a collection of simple directed paths and cycles that cover all edges of $\gG$ while respecting the capacity constraints. Define the source and sink vertex sets as
\begin{equation*}
    \gS:=\{v\in \gV \mid \outde (v)>\inde (v) \},\qquad \gD:=\{v\in \gV \mid \outde (v)<\inde (v) \},
\end{equation*}
and call vertices with $\outde(v)=\inde(v)$ balanced. If every vertex in the graph is balanced, then $\gG$ is an Eulerian digraph, meaning that its edges can be decomposed entirely into directed cycles. Our approach is therefore to first find a set of simple directed paths that originate in $\gS$ and terminate in $\gD$. Removing these paths from $\gG$ yields a residual subgraph in which all vertices are balanced, and hence the residual is Eulerian.


To extract a collection of simple directed paths, we interpret the directed edges of $G$ as arcs of a unit-capacity flow network.  
Construct a flow network $\gG'=(\gV',\gE')$ as follows:  
retain all original vertices $\gV$ and edges $\gE$, and assign to each original edge $e\in \gE$ a capacity $c(e)=1$.  
Add two additional vertices, a super-source $\mathbf{s}$ and a super-sink $\mathbf{t}$.  
For each $v\in \gS$, add an arc $(\mathbf{s},v)$ with capacity $c(\mathbf{s},v)=\outde(v)-\inde(v)$;  
for each $v\in \gD$, add an arc $(v,\mathbf{t})$ with capacity $c(v,\mathbf{t})=\inde(v)-\outde(v)$.  Figure~\ref{fig:flow-graph} includes a constructed flow network.
Since $\sum_{v\in V} \outde(v)-\inde(v)=0$, we have 
\[
\sum_{v\in \gS} \outde(v)-\inde(v) = \sum_{v\in \gD} \inde(v)-\outde(v):=C',
\]
so the total supply out of $\mathbf{s}$ equals the total demand into $\mathbf{t}$. Therefore, the max flow on $\gG'$ from $\mathbf{s}$ to $\mathbf{t}$ is with capacity $C'$. Because all original edges have unit capacity, no original edge is used more than once when carrying flow.  
Furthermore, standard augmenting-path algorithms (e.g., Edmonds--Karp algorithm \citealt{edmonds1972theoretical}, see also Algorithm~\ref{alg:edmonds-karp} in the appendix) ensure that every augmenting path encountered during the construction is simple in $\gG'$, and therefore the projected path in $\gG$ is also simple. For simplicity, we define the mapping $\gF^{-1}$ which projects the path $\path'$ on $\gG'$ back to the original path $\path$ on $\gG$. Thus the support of the max flow on original edges decomposes into edge-disjoint simple directed paths starting at vertices in $\gS$ and ending at vertices in $\gD$. 
Let $\widetilde{\gP}_p$ denote this family of paths on the auxiliary unbalanced directed graph.

Remove from $\gG$ all edges used by the paths in $\mathcal{P}$ along with the nodes that subsequently have both in-degree and out-degree equal to 0. Let the resultingdirected graph be denoted by $\gG_{\mathrm{rem}}=(\gV_{\mathrm{rem}},\gE_{\mathrm{rem}})$.  
For each $v\in \gS$, exactly $\outde(v)-\inde(v)$ outgoing edges are removed and no incoming edges are removed. For each $v\in \gD$, exactly $\inde(v)-\outde(v)$ incoming edges are removed and no outgoing edges are removed. Moreover, for each originally balanced vertex $v$, the removal of a path that passes through $v$ eliminates one incoming and one outgoing edge leaving the vertex still balanced. Hence for all $v\in \gV_{\mathrm{rem}}$, the remaining in-degree and out-degree satisfy $\deg^+_{\gG_{\mathrm{rem}}}(v) = \deg^-_{\gG_{\mathrm{rem}}}(v)$.
Thus, $\gG_{\mathrm{rem}}$ is an Eulerian directed graph (possibly disconnected). A classical graph-theoretic result (see, e.g., \citealt{bang2008digraphs}) states
that every Eulerian directed graph decomposes into edge-disjoint simple directed cycles. For completeness, we present an explicit cycle-decomposition procedure in Algorithm~\ref{alg:eulerian-cycle-decomposition} in the appendix. Applying this result to $\gG_{\mathrm{rem}}$ yields a family $\widetilde{\gP}_c$ of simple directed cycles whose edges partition $\gE_{\mathrm{rem}}$.

Finally, projecting each path in $\widetilde{\gP}_p$ and each cycle in $\widetilde{\gP}_c$ back by $\gA^{-1}$ can offer us a corresponding collection of alternating paths and alternating cycles in the original disagreement set. The resulting set of the alternating paths and cycles satisfies all the conditions in Theorem \ref{thm:colloction-path}. The details of the whole procedure can be found in Algorithm \ref{alg:path-cycle-decomposition-highlevel}. Once such a decomposition is obtained, we can directly apply the AP design and all inference guarantees developed for the one-to-one matching case.

\begin{algorithm}[t]
  \caption{Alternating Paths and Cycles Decomposition}
  \label{alg:path-cycle-decomposition-highlevel}
  \DontPrintSemicolon
  \begin{algorithmic}[1]
    \STATE \textbf{Input:} Auxiliary Unbalanced Directed graph $\gG = (\gV,\gE)$ of Disagreement set $\Diff$ and the mapping $\gA^{-1}$
    \STATE \textbf{Output:} The set of alternating paths and cycles $\Path$

    \STATE Compute $\outde(v)$ and $\inde(v)$ for all $v \in \gV$, and set $b(v) \gets \outde(v) - \inde(v)$
    \STATE $\gS \gets \{ v \in \gV : b(v) > 0 \}$ and $\gD \gets \{ v \in \gV : b(v) < 0 \}$

    \STATE Construct flow network $\gG'=(\gV',\gE')$ by:
    \STATE \quad adding super-source $\mathbf{s}$ and super-sink $\mathbf{t}$,
    \STATE \quad keeping all edges $e \in \gE$ with capacity $c(e) = 1$,
    \STATE \quad adding edges $(\mathbf{s},v)$ with capacity $b(v)$ for all $v \in \gS$,
    \STATE \quad adding edges $(v,\mathbf{t})$ with capacity $-b(v)$ for all $v \in \gD$.
\STATE $\widetilde{\gP}_p' \gets \textsc{Edmonds--Karp}(\gG')$
    \STATE Let $\widetilde{\gP}_p$ be the set of edge-disjoint simple paths induced by $\widetilde{\gP}_p'$ on the original edges $\gE$

    \STATE Remove from $\gG$ all edges belonging to paths in $\mathcal{P}$ and vertices with zero degree, and denote the remaining graph by $\gG_{\mathrm{rem}}=(\gV_{\mathrm{rem}},\gE_{\mathrm{rem}})$
    \STATE $\widetilde{\gP}_c \gets \textsc{EulerianCycleDecomposition}(\gG_{\mathrm{rem}})$

    \STATE \textbf{return} $\{\gA^{-1}(\path),\; \forall \path \in \widetilde{\gP}_p \cup \widetilde{\gP}_c\}$
  \end{algorithmic}
\end{algorithm}

\section{Numerical Results}
We evaluate the real-world performance of our AP design on one of the largest workforce datasets provided by Revelio Lab\footnotemark
\footnotetext{See, \url{https://wrds-www.wharton.upenn.edu/pages/about/data-vendors/revelio-labs/}.}. The dataset contains rich job–worker matching information across a broad range of occupations and industries. Following \cite{tang2024match}, we focus on a subset consisting of mid-level software engineers in the United States during the period 2010–2015, with employment duration as the primary outcome of interest.

To construct a controlled experimental environment, we cluster workers and jobs into 50 groups each. Using the full employment history data, we form a ground-truth matching-value matrix $\mathbf{Y}^*$ where each entry is defined as the average of an indicator for whether an employment spell exceeds six months, taken over all worker–job matches between worker group $i$ and job group $j$. With full access to $\mathbf{Y}^*$, the entry $Y^*_{i,j}$ can be interpreted as the probability that a worker from group $i$ remains employed for longer than six months when matched to a job from group $j$. By construction, $Y^*_{i,j}\in[0,1]$. Throughout the numerical study, we interpret this probability as a proxy for worker satisfaction with the assigned job: a higher value of $Y^*_{i,j}\in[0,1]$ indicates a more stable and hence more desirable match from the worker’s perspective.


Table~\ref{tab:variance_comparison} reports a variance comparison in a controlled setting with $n$ workers and $n$ jobs, where one worker and one job are drawn from each group and $n$ ranges from 10 to 50. The two matching plans under comparison are constructed as follows. Under the treatment plan, worker $k$ is matched to job $k$. Under the control plan, worker $k$ is matched to job $k+1$ for $1\le k\le n-1$, and worker $n$ is matched to job $1$, forming a cyclic shift. The table compares the AP design (with randomization probability $p=0.5$) against the naive design described in Remark~\ref{remark:naive}. ``True Var” denotes the true variance of the estimator computed using the full matching-value matrix $\mathbf{Y}^*$. ``Empirical Var” is the sample variance obtained from 500 independent randomizations. ``Var UB” refers to the theoretical variance upper bound, and ``Est. Var UB” denotes its data-driven estimate derived from Proposition~\ref{prop:estimated-variance}. Consistent with our theoretical analysis, the naive design exhibits a variance that does not diminish as $n$ increases. In contrast, the variance of the AP design decreases steadily with the size of the problem, illustrating its scalability. Moreover, the estimated variance upper bound closely tracks the true variance across all values of $n$, suggesting that the proposed bound is both informative and practically tight.


\begin{table}[htbp]
\centering
\caption{Variance and Bias Comparison: AP Design vs Naive Estimator.}
\label{tab:variance_comparison}
\begin{tabular}{c *{8}{c}}
\toprule
Size $n$ & \multicolumn{5}{c}{AP Design} & \multicolumn{3}{c}{Naive Design}  \\
\cmidrule(lr){2-6}  \cmidrule(lr){7-9}
& True Var & Emp. Var & Var UB & Est. Var UB & Bias & True Var & Emp. Var & Bias \\
\midrule
10 & 0.8983 & 0.9511 & 0.9024 & 0.9173 & 0.0128 & 3.1155 & 3.1217 & 0.0071 \\
20 & 0.4475 & 0.4564 & 0.4506 & 0.4479 & 0.0074 & 3.0317 & 3.0142 & -0.1532 \\
30 & 0.3075 & 0.3644 & 0.3090 & 0.3114 & 0.0223 & 3.1029 & 3.0872 & -0.1480 \\
40 & 0.2311 & 0.2102 & 0.2321 & 0.2321 & 0.0248 & 3.1011 & 3.1024 & -0.0704 \\
50 & 0.1848 & 0.1920 & 0.1855 & 0.1859 & -0.0122 & 3.0925 & 3.0916 & -0.0844 \\
\bottomrule
\end{tabular}
\footnotesize

\vspace{0.2cm}
Note: ``Ture Var" is the true variance of the estimator. ``Emp. Var'' stands for empirical variance computed from 500 simulation runs. ``Var UB" is the variance upper bound derived in Proposition~\ref{prop:estimated-variance}. ``Est. Var UB" is the average estimated variance upper bound by Proposition~\ref{prop:estimated-variance} over 500 simulation runs.
\end{table}

\begin{figure}[h]
		\centering
		\hskip -0.075in
		\begin{minipage}[t]{0.5\linewidth}
			    \includegraphics[width=1\textwidth]{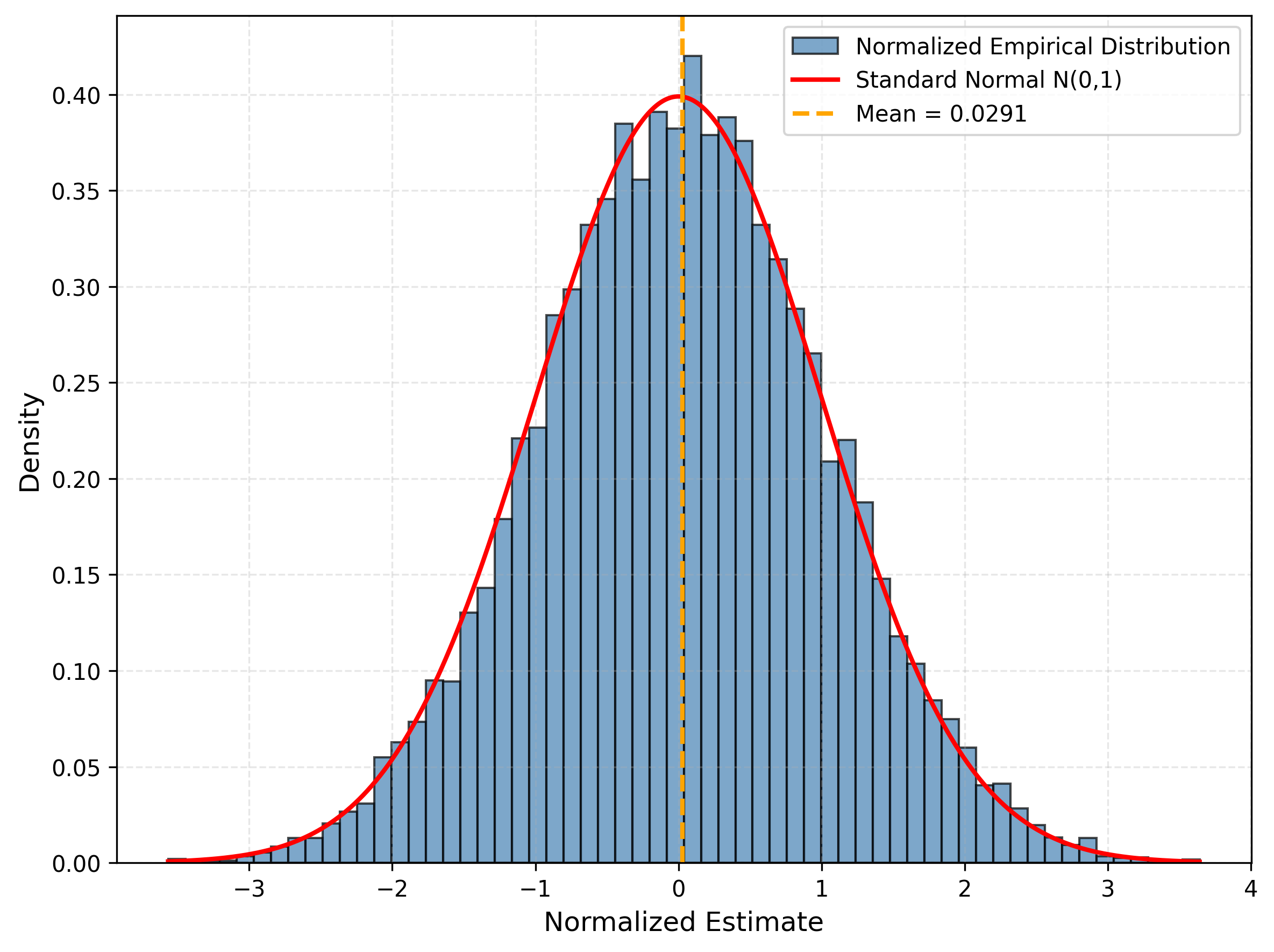}
    \caption{Normalized empirical distribution}
    \label{fig:empirical-distribution}
		\end{minipage}
		\begin{minipage}[t]{0.5\linewidth}
			\includegraphics[width=1\textwidth]{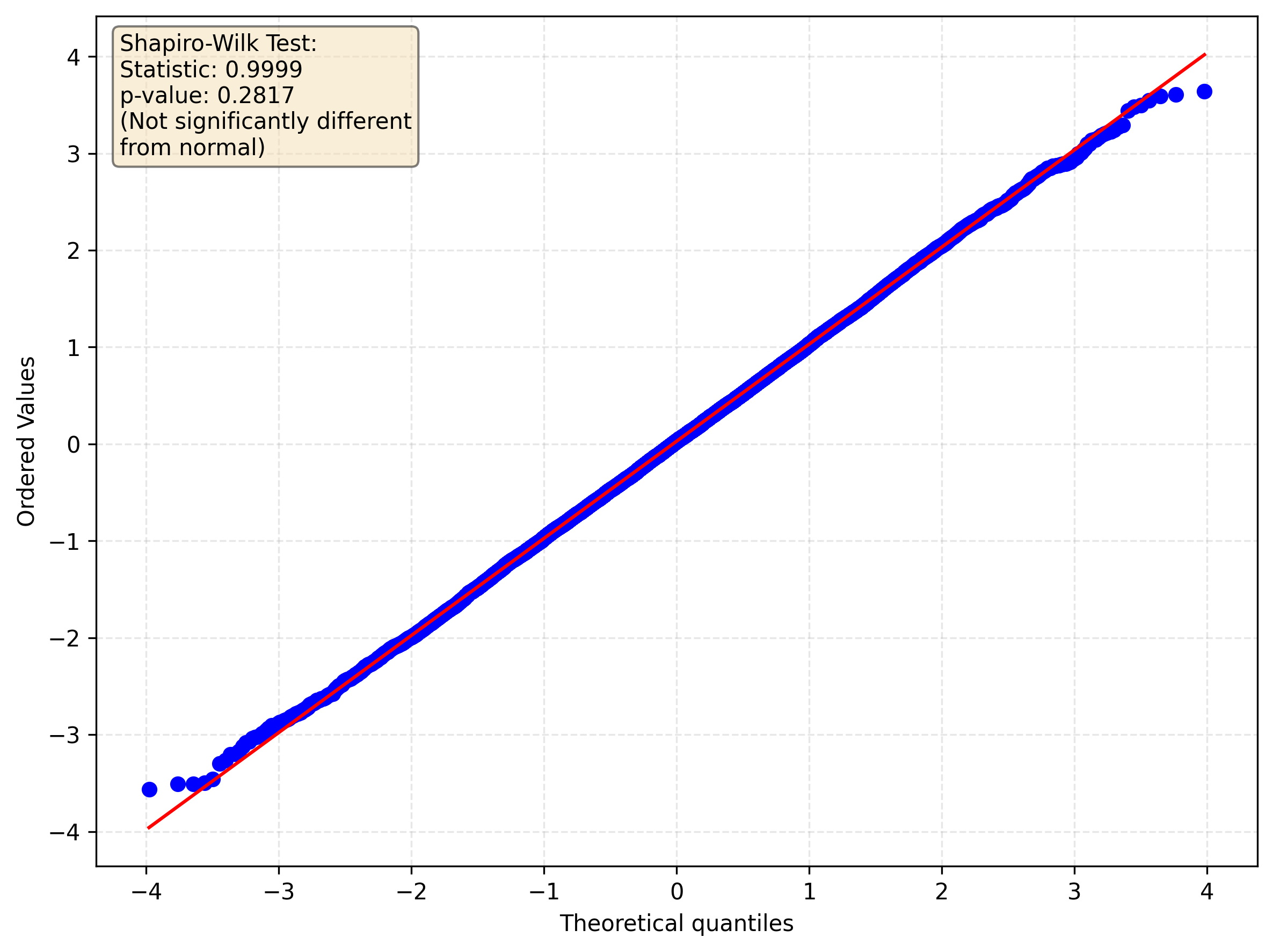}
				\caption{Q-Q Plot against the standard normal}
				\label{fig:QQ-plot}
		\end{minipage}	
	\end{figure}

To verify the central limit theorem, we plot out the normalized empirical distribution of 5000 independent runs of the randomization in Figure~\ref{fig:empirical-distribution}, together with a standard normal as a comparison. Furthermore, Figure~\ref{fig:QQ-plot} presents a Q–Q plot comparing the empirical quantiles of the sample with those of a standard normal distribution. The points align closely with the 45-degree reference line across the entire range, indicating that the empirical distribution exhibits strong agreement with normality. Only negligible deviations are observed in the extreme tails. This visual evidence is consistent with the results of the Shapiro–Wilk test (test statistic $= 0.9999$, p-value $= 0.2817$), which fails to reject the null hypothesis of normality. Collectively, these findings indicate that the distribution of the data does not significantly deviate from a standard normal distribution.

\begin{figure}[h]
        \centering
        \includegraphics[width=0.8\linewidth]{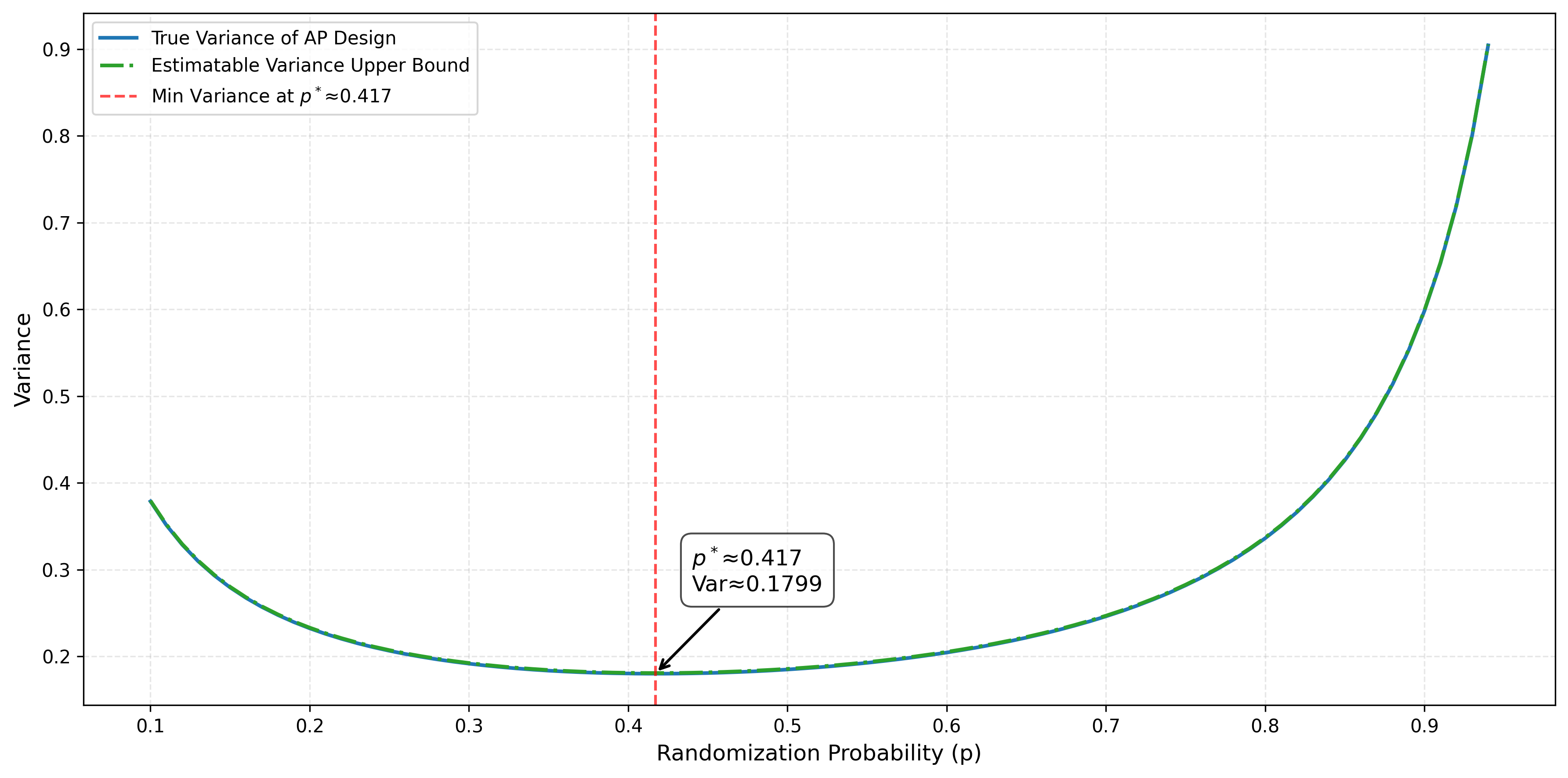}
        \caption{Variance of AP design under different randomization probability $p$}
        \label{fig:empirical-optimal-p}
    \end{figure}

   Figure~\ref{fig:empirical-optimal-p} illustrates how the variance of the AP estimator and its estimable upper bound from Proposition~\ref{prop:estimated-variance} vary with the randomization probability $p$ when $n=50$. The true variance and the upper bound nearly coincide across the entire range of $p$, providing further empirical evidence that the proposed variance bound is tight for this dataset. The empirically optimal randomization probability is $p^*\simeq 0.417$, which is close to the worst-case optimal value $\sqrt{2}-1$. An additional observation is that the variance curve is relatively flat in a neighborhood around $p^*$, indicating that the performance of the AP design is robust to moderate misspecification of the optimized randomization probability.

\section{Discussion and Concluding Remarks}
In this work, we study a fundamental yet underexplored challenge in operations management: how to experimentally compare matching mechanisms in an assumption-free, design-based manner in the presence of matching interference. While much of the existing literature focuses on mitigating bias or improving efficiency within a fixed matching mechanism, we take a complementary perspective and develop a principled framework for directly testing two predetermined matching plans, such as a newly proposed algorithm against a status quo, without imposing outcome or behavioral models.

Our key methodological contribution is the Alternating Path Randomized Design, which exploits the structure of the disagreement set between two matchings. In the one-to-one setting, we show that this disagreement set admits a unique decomposition into alternating paths and cycles, along which sequential randomization can be conducted while respecting feasibility constraints. Within a minimax framework, we characterize the variance properties of the resulting Horvitz–Thompson estimator and show that, for long paths and cycles, the optimal conditional randomization probability converges to $\sqrt{2}-1$. We further establish a finite-population central limit theorem that remains valid even when the configuration of paths and cycles is complex, heterogeneous, and potentially unstable as the population grows.

We extend the framework to many-to-one matchings, where capacity constraints introduce substantial new challenges. By mapping the disagreement structure to an auxiliary unbalanced directed graph, we reduce the construction of feasible alternating paths and cycles to classical graph-theoretic problems involving augmenting paths and Euler-tour decompositions. This extension highlights that adapting experimental designs from one-to-one to many-to-one settings is far from mechanical and requires careful attention to feasibility and interference.

Overall, this work offers a unified, design-based methodology for testing matching mechanisms under interference, with direct relevance for platforms and policymakers seeking to validate new matching algorithms before full-scale deployment. We hope these results will encourage further research on experimental design in complex allocation systems where interference is intrinsic rather than incidental.


	\bibliographystyle{informs2014}
	\bibliography{ref.bib}
 	\clearpage 
	
	\ECSwitch
	

	\begin{center}
		\large{Online Appendix for ``Experimental Design for Matching''}
	\end{center}

\section{Technical Tools}

The first important technical tool is the non-stationary $\alpha$-Mixing CLT by \cite{bradley2017central} in their Theorem 1.1.
\begin{theorem}[Non-Stationary $\alpha$-Mixing CLT, \citealt{bradley2017central}] \label{thm:non-stationary-CLT}
Suppose \(d\) is a positive integer. For each \(n \in \mathbb{N}\), suppose \(L_n := (L_{n1}, L_{n2}, \dots, L_{nd})\) is an element of \(\mathbb{N}^d\), and suppose \(X^{(n)} := (X^{(n)}_k, k \in B(L_n))\) is an array of random variables such that for each \(k \in B(L_n)\), \(\mathbb{E} X^{(n)}_k = 0\) and \(\mathbb{E}(X^{(n)}_k)^2 < \infty\). Suppose the following mixing assumptions hold:
\[
\alpha(m) := \sup_n \alpha(X^{(n)}, m) \to 0 \quad \text{as } m \to \infty,
\]
\[
\rho'(1) := \sup_n \rho'(X^{(n)}, 1) < 1.
\]
For each \(n \in \mathbb{N}\), define the random sum \(S(X^{(n)}, L_n) = \sum_{k \in B(L_n)} X^{(n)}_k\), define the quantity \(\sigma_n^2 := \mathbb{E}[S(X^{(n)}, L_n)]^2\), and assume that \(\sigma_n^2 > 0\). Suppose also that the Lindeberg condition
\[
\forall \varepsilon > 0, \quad \lim_{n \to \infty} \frac{1}{\sigma_n^2} \sum_{k \in B(L_n)} \mathbb{E}[(X^{(n)}_k)^2 \sI(|X^{(n)}_k| > \varepsilon \sigma_n)] = 0
\]
holds. Then \(\sigma_n^{-1} S(X^{(n)}, L_n) \xrightarrow{d} N(0, 1)\) as \(n \to \infty\), where \(\xrightarrow{d}\) denotes convergence in distribution.
\end{theorem}

The theorem relies on the following concepts defined on a probability space \((\Omega, \mathcal{F}, P)\):

\begin{itemize}
\item \textbf{Strong Mixing Coefficient}: For any two \(\sigma\)-fields \(A, B \subseteq \mathcal{F}\),
\[
\alpha(A, B) := \sup_{A \in A, B \in B} |P(A \cap B) - P(A)P(B)|.
\]
\item \textbf{Maximal Correlation Coefficient}: For any two \(\sigma\)-fields \(A, B \subseteq \mathcal{F}\),
\[
\rho(A, B) := \sup \{ |\Corr(f, g)| : f \in L^2_{\text{real}}(A), g \in L^2_{\text{real}}(B) \}.
\]
\item \textbf{Random Field}: \(X := (X_k, k \in \mathbb{Z}^d)\) is a random field, not necessarily stationary.
\item \textbf{Box \(B(L)\)}: For \(L = (L_1, L_2, \dots, L_d) \in \mathbb{N}^d\),
\[
B(L) := \{ k = (k_1, k_2, \dots, k_d) \in \mathbb{N}^d : \forall u \in \{1, 2, \dots, d\}, 1 \leq k_u \leq L_u \},
\]
with \(|B(L)| = L_1 \cdot L_2 \cdot \dots \cdot L_d\).
\item \textbf{\(\alpha\)-Mixing for Random Fields}: The random field \(X\) is \(\alpha\)-mixing if
\[
\alpha(X, n) := \sup \alpha(\sigma(X_k, k \in Q), \sigma(X_k, k \in S)),
\]
where the supremum is over all pairs of nonempty, disjoint sets \(Q, S \subset \mathbb{Z}^d\) such that there exist \(u \in \{1, 2, \dots, d\}\) and \(j \in \mathbb{Z}\) with \(Q \subset \{ k : k_u \leq j \}\) and \(S \subset \{ k : k_u \geq j + n \}\), and \(\alpha(X, n) \to 0\) as \(n \to \infty\).
\item \textbf{\(\rho'\) Coefficient}: The coefficient \(\rho'(X^{(n)}, 1)\) is the supremum of \(\rho(A, B)\) over pairs of nonempty, disjoint sets \(Q, S \subset B(L_n)\) such that there exist \(u \in \{1, 2, \dots, d\}\) and \(j \in \mathbb{Z}\) with \(Q \subset \{ k : k_u \leq j \}\) and \(S \subset \{ k : k_u \geq j + 1 \}\).
\end{itemize}

\begin{theorem}[Lindeberg-Feller Central Limit Theorem, \citealt{feller1991introduction}]
Consider a triangular array of random variables \(\{X_{n,k} : n \in \mathbb{N}, k = 1, \dots, n\}\), where for each fixed \(n\), the variables \(X_{n,1}, \dots, X_{n,n}\) are independent, with \(\mathbb{E}[X_{n,k}] = 0\) and \(\mathrm{Var}(X_{n,k}) = \sigma_{n,k}^2 < \infty\). Let \(s_n^2 = \sum_{k=1}^n \sigma_{n,k}^2\), and assume \(s_n^2 > 0\). Suppose the Lindeberg condition holds: for every \(\varepsilon > 0\),
\[
\lim_{n \to \infty} \frac{1}{s_n^2} \sum_{k=1}^n \mathbb{E}\left[ X_{n,k}^2 \sI(\{|X_{n,k}| > \varepsilon s_n\}) \right] = 0.
\]
Then, as \(n \to \infty\),
\[
\frac{1}{s_n} \sum_{k=1}^n X_{n,k} \xrightarrow{d} \mathcal{N}(0,1),
\]
where \(\xrightarrow{d}\) denotes convergence in distribution.
\end{theorem}

The next technical theorem we are relying on is Martingale Rosenthal Inequality, which is the Thereom 6.6.7 in \cite{de1999decoupling}.
\begin{theorem}[Martingale Rosenthal Inequality, \citealt{de1999decoupling}] \label{thm:rosenthal}
Let $p \ge 2$. Let $M_n = \sum_{i=1}^n d_i$ where $\{M_n, \mathcal{F}_n\}$ is a martingale
with martingale difference sequence $\{d_i\}$. Set $\|X\|_p = (\mathbb{E}|X|^p)^{1/p}$ for any random
variable $X$, and define
\[
U_{n,p} = \left( \sum_{j=1}^n \|d_j\|_p^p \right)^{1/p}, \qquad
s_n(M) = \left( \sum_{i=1}^n \mathbb{E}(|d_i|^2 \mid \mathcal{F}_{i-1}) \right)^{1/2}.
\]
Then there exist constants $0 < c_p, C_p < \infty$ such that
\[
c_p \big[ \| s_n(M) \|_p^p + U_{n,p}^p \big]
   \le \| M_n \|_p^p
   \le C_p \big[ \| s_n(M) \|_p^p + U_{n,p}^p \big].
\]
\end{theorem}

Another useful form of Bernstein inequality for reversible Markov Chain by \cite{Paulin2012ConcentrationIF} is as following.
\begin{theorem}[Bernstein Inequality, Theorem 3.9 by \cite{Paulin2012ConcentrationIF}]\label{thm:bernstein_markov}
Let $X_1, \ldots, X_n$ be a stationary reversible Markov chain with finite state space~$\Omega$, absolute spectral gap~$\gamma^*$ and the stationary distribution $\pi$. 
Let $f_1, f_2, \ldots, f_n \in L^2(\pi)$ be functions satisfying 
$\lvert f_i(x) - \mathbb{E}_\pi(f_i)\rvert \le C$ for every $x \in \Omega$. 
Let 
\[
S' := \sum_{i=1}^n f_i(X_i)
\quad\text{and}\quad 
V_{S'} := \sum_{i=1}^n \Var_\pi(f_i).
\]
Then for every $n \ge 1$ and $t \ge 0$, we have
\begin{equation}\label{eq:bernstein_reversible}
    \mathbb{P}_\pi\!\left( \bigl|S' - \mathbb{E}_\pi(S')\bigr| \ge t \right)
    \le 2 \exp\!\left(
    -\,\frac{t^2 \cdot \bigl(2\gamma^* - (\gamma^*)^2\bigr)}{8V_{S'} + 20tC}
    \right).
\end{equation}
\end{theorem}

The following theorem is useful to bounded the fourth moment for a strong mixing sequence established by \cite{rio2017asymptotic} in their Theorem 2.1.
\begin{theorem}[Fourth–moment bound under strong mixing {\citealt{rio2017asymptotic}}]\label{thm:fourth-moment}
Let $\{X_k\}_{k\ge1}$ be a sequence of centered real–valued random variables with
$\mathbb E|X_k|^4<\infty$ for all $k$, and put $S_n:=\sum_{k=1}^n X_k$.
Let $\alpha(m)$ denote the strong mixing coefficients of $\{X_k\}_{k\ge 1}$ and
\[
\alpha^{-1}(u)\ :=\ \inf\{\,m\ge 0:\ \alpha(m)\le u\,\},\qquad u\in(0,1].
\]
For each $k$, let $Q_k$ be the quantile function of $|X_k|$, i.e.
\[
Q_k(u)\ :=\ \inf\{\,t\ge 0:\ \mathbb P(|X_k|>t)\le u\,\},\qquad u\in(0,1].
\]
Define
\[
M_{4,\alpha,n}\big(Q_k\big)\ :=\ \sum_{k=1}^n \int_0^1
\big[\alpha^{-1}(u)\wedge n\big]^3\, Q_k^4(u)\,du .
\]
Then, for every $n\ge1$,
\[
\mathbb E\big(S_n^4\big)\ \le\
3\Bigg(\sum_{i=1}^n\sum_{j=1}^n \big|\mathbb E(X_i X_j)\big|\Bigg)^{\!2}
\;+\;48\, M_{4,\alpha,n}\big(Q_k\big).
\]
\end{theorem}

\section{Algorithm Details}

In this section, Algorithm \ref{alg:edmonds-karp} shows the details of the classical Edmonds--Karp used in Algorithm \ref{alg:path-cycle-decomposition-highlevel}. Algorithm \ref{alg:edmonds-karp} relies on another standard routine of breath-first search, which is presented in Algorithm~\ref{alg:bfs-residual}. Algortihm~\ref{alg:eulerian-cycle-decomposition} is the routine of Eulerian Cycle Decomposition. 

\begin{algorithm}[t]
\caption{Edmonds--Karp}
\label{alg:edmonds-karp}
\begin{algorithmic}[1]
  \STATE \textbf{Input:} Directed graph $G=(V,E)$ with capacities $c(u,v)$, source $s$, sink $t$
  \STATE \textbf{Output:} List $\mathcal{A}$ of all augmenting $s$--$t$ paths discovered during the algorithm
  \STATE Initialize residual capacities $r(u,v) \gets c(u,v)$ for all edges $(u,v)$
  \STATE $\mathcal{A} \gets \emptyset$
  \WHILE{true}
    \STATE $P \gets \textsc{BFS-Residual}(G,r,s,t)$
    \IF{$P = \textsc{none}$}
      \STATE \textbf{break}
    \ENDIF
    \STATE  $\mathcal{A} \gets \mathcal{A} \cup P$
    \STATE $\Delta \gets \min\{ r(u,v) : (u,v) \in P \}$
    \FORALL{edges $(u,v)$ in $P$}
      \STATE $r(u,v) \gets r(u,v) - \Delta$
    \ENDFOR
  \ENDWHILE
  \STATE \textbf{return} $\mathcal{A}$
\end{algorithmic}
\end{algorithm}

\begin{algorithm}[t]
\caption{\textsc{BFS-Residual}$(G,r,s,t)$}
\label{alg:bfs-residual}
\begin{algorithmic}[1]
  \STATE \textbf{Input:} Residual graph $(G,r)$ with residual capacities $r(u,v)$, source $s$, sink $t$
  \STATE \textbf{Output:} Augmenting path $P$ from $s$ to $t$, or \textsc{none} if no such path exists
  \STATE \textbf{Initialization:} Mark all vertices as unvisited, $\mathrm{parent}(v) \gets \textsc{none}$ for all $v \in V(G)$, create an empty queue $Q$
  \STATE Enqueue $s$ into $Q$ and mark $s$ as visited
  \WHILE{$Q$ is not empty}
    \STATE Dequeue $u$ from $Q$
    \FORALL{edges $(u,v)$ with $r(u,v) > 0$}
      \IF{$v$ is unvisited}
        \STATE Mark $v$ as visited, $\mathrm{parent}(v) \gets u$, and enqueue $v$ into $Q$
        \IF{$v = t$}
          \STATE \textbf{return} $P$ by back tracking the parent node till $s$.
        \ENDIF
      \ENDIF
    \ENDFOR
  \ENDWHILE
  \STATE \textbf{return} \textsc{none}
\end{algorithmic}
\end{algorithm}

\begin{algorithm}[t]
\caption{\textsc{EulerianCycleDecomposition}$(G_{\mathrm{rem}})$}
\label{alg:eulerian-cycle-decomposition}
\begin{algorithmic}[1]
  \STATE \textbf{Input:} Eulerian directed graph $G_{\mathrm{rem}} = (V_{\mathrm{rem}}, E_{\mathrm{rem}})$ 
  \STATE \textbf{Output:} Family $\mathcal{C}$ of simple directed cycles whose edges partition $E_{\mathrm{rem}}$
  \STATE \textbf{Initialization:} For each edge $e \in E_{\mathrm{rem}}$, mark $e$ as unused,  $\mathcal{C} \gets \emptyset$ and build adjacency lists $\mathrm{Adj}[v]$ of outgoing neighbors for all $v \in V$
  \WHILE{there exists an unused edge in $E_{\mathrm{rem}}$}
    \STATE Choose a vertex $v_0$ incident to some unused edge
    \STATE \COMMENT{Step 1: build a closed directed walk}
    \STATE Initialize empty list $W$, $v \gets v_0$
    \STATE Append $v$ to $W$
    \WHILE{there exists an unused outgoing edge $(v,u)$ from $v$}
      \STATE Choose such an edge $(v,u)$ and mark edge $(v,u)$ as used
      \STATE $v \gets u$ and append $v$ to $W$
    \ENDWHILE
    
    \STATE \COMMENT{Step 2: extract one or more simple directed cycles from $W$}
    \STATE Initialize a map $\mathrm{firstPos}(x) \gets -1$ for all $x \in V_{\mathrm{rem}}$ and $i \gets 0$
    \WHILE{$i < |W|$}
      \STATE Let $x \gets W[i]$
      \IF{$\mathrm{firstPos}(x) = -1$}
        \STATE $\mathrm{firstPos}(x) \gets i$ and $i \gets i+1$
      \ELSE
        \STATE $j \gets \mathrm{firstPos}(x)$
        \STATE Define cycle $C$ as the sequence of vertices $W[j], W[j+1], \ldots, W[i]$
        \STATE $\mathcal{C}\gets \mathcal{C} \cup \{C\}$
        \STATE \COMMENT{Allow vertices between $j$ and $i$ to participate in other cycles}
        \FOR{$k \gets j+1$ to $i-1$}
          \STATE $\mathrm{firstPos}(W[k]) \gets -1$
        \ENDFOR
        \STATE $\mathrm{firstPos}(x) \gets i$ and $i \gets i+1$
      \ENDIF
    \ENDWHILE
  \ENDWHILE
  \STATE \textbf{return} $\mathcal{C}$
\end{algorithmic}
\end{algorithm}

\section{Proof to Lemma \ref{lemma:uncondition-prob}}
\begin{proof}{Proof to Lemma \ref{lemma:uncondition-prob}.}
    We first consider the normal case where $\path_i$ is not a cycle or $j\not= k(i)$. In this case, for $j>1$, we can have
    \begin{align*}
        \prob(W_{i,j}=1)&=\prob(W_{i,j}=1, W_{i,j-1}=0)\\
        &=\prob(W_{i,j}=1\mid W_{i,j-1}=0)\cdot \prob(W_{i,j-1}=0)\\
        &=p\cdot(1-\prob(W_{i,j-1}=1)),
    \end{align*}
    where the first equation due to the fact that $W_{i,j}=1$ implies $W_{i,j-1}=0$, and the last equation holds according to AP design. By simple calculation, we can know that $\prob(W_{i,j}=1)-\frac{p}{p+1}=-p\cdot (\prob(W_{i,j-1}=1)-\frac{p}{p+1})$ allowing us to use the recursion. With the initial value $\prob(W_{i,1}=1)=\frac{p}{1+p}$, we have
     \begin{equation*}
       \prob(W_{i,j}=1)=\frac{p}{p+1}.
    \end{equation*}

    For the last edge of a cycle $\path_i$ (i.e., $j=k(i)$), we can have
        \begin{align}
        \prob(W_{i,j}=1)&=\prob(W_{i,j}=1, W_{i,j-1}=0,W_{i,1}=0)\notag\\
        &=\prob(W_{i,j}=1\mid W_{i,j-1}=0,W_{i,1}=0)\cdot \prob(W_{i,j-1}=0\mid W_{i,1}=0)\cdot \prob( W_{i,1}=0)\notag \\
        &=(1-\frac{p}{p+1})\cdot\prob(W_{i,j-1}=0\mid W_{i,1}=0). \label{eq:cycle-noncondition-prob}
    \end{align}
    Now we only need to calculate $\prob(W_{i,j-1}=0\mid W_{i,1}=0)$.
\begin{align*}
    \prob(W_{i,j-1}=0\mid W_{i,1}=0)&=1- \prob(W_{i,j-1}=1\mid W_{i,1}=0)\\
    &=1- \prob(W_{i,j-1}=1,W_{i,j-2}=0\mid W_{i,1}=0)\\
    &=1- \prob(W_{i,j-1}=1\mid W_{i,j-2}=0, W_{i,1}=0)\prob(W_{i,j-2}=0\mid W_{i,1}=0)\\
    &=1- p\cdot \prob(W_{i,j-2}=0\mid W_{i,1}=0).
\end{align*}
Again, by recursion, we can have
\begin{equation}\label{eq:zero-to-zero}
    \prob(W_{i,j-1}=0\mid W_{i,1}=0)-\frac{1}{p+1}=(-p)^{j-2} \left(\prob(W_{i,1}=0\mid W_{i,1}=0)-\frac{1}{p+1}\right)=-\frac{(-p)^{j-1}}{p+1},
\end{equation}
which leads to $\prob(W_{i,j-1}=0\mid W_{i,1}=0)=\frac{1}{p+1}-\frac{(-p)^{j-1}}{p+1}$. By plugging this result back to  \Eqref{eq:cycle-noncondition-prob}, we finish the prove. 
A side note here is that when $j=k(i)$ and $\path_i$ is a cycle, $j$ will always be larger than 4 and thus $j-2$ in the above equations is well defined. \qed
\end{proof}

\section{Proof to Theorem \ref{thm:variance-decompose}}

\begin{proof}{Proof to Theorem \ref{thm:variance-decompose}}
     A more compact way to write $\EPTE_i$ is
    \begin{equation}\label{eq:compact-estimator}
        \EPTE_i=\sum_{j=1}^{k(i)}(-1)^{j+1}\frac{\sI\{w_{i,j}=1\}\cdot Y_{v_{i,j},v_{i,j+1}}}{\prob(W_{i,j}=1)}.
    \end{equation}
For simplicity, we introduce the notation $\tilde{Y}_{v_{i,j},v_{i,j+1}}$ to denote $\frac{\sI\{w_{i,j}=1\}\cdot Y_{v_{i,j},v_{i,j+1}}}{\prob(W_{i,j}=1)}$. Based on \Eqref{eq:compact-estimator}, the variance of $\EPTE_i$ can be decomposed as follows,
\begin{align}
    \Var(\EPTE_i) &= \Var\left(\sum_{j=1}^{k(i)}(-1)^{j+1} \tilde{Y}_{v_{i,j},v_{i,j+1}} \right)\notag\\
    &=\sum_{j=1}^{k(i)} \Var\left(\tilde{Y}_{v_{i,j},v_{i,j+1}}\right)+2 \sum_{0\le j<q \le k(i)} (-1)^{j+q+2} \Cov\left(\tilde{Y}_{v_{i,j},v_{i,j+1}},\tilde{Y}_{v_{i,q},v_{i,q+1}}\right). \label{eq:variance-decompose}
\end{align}

We start with the simpler case where $\path_i$ is a path.
For the first term in \Eqref{eq:variance-decompose}, we can have
\begin{align}  \Var\left(\tilde{Y}_{v_{i,j},v_{i,j+1}}\right)&=\frac{Y_{v_{i,j},v_{i,j+1}}^2}{(\prob(W_{i,j}=1))^2} \Var(\sI\{w_{i,j}=1\})\notag\\
    &=\frac{1-\prob(W_{i,j}=1)}{\prob(W_{i,j}=1)} Y_{v_{i,j},v_{i,j+1}}^2\notag\\
    &=\frac{1}{p}Y_{v_{i,j},v_{i,j+1}}^2,\label{eq:variance-single-term}
\end{align}
where the second equation holds based on the variance of a binary distribution, and plugging in \Eqref{eq:uncondition-prob} leads to the last equation. For the second term in \Eqref{eq:variance-decompose}, we have the follows.
\begin{align}
\Cov\left(\tilde{Y}_{v_{i,j},v_{i,j+1}},\tilde{Y}_{v_{i,q},v_{i,q+1}}\right)&= \E\left[\frac{\sI\{W_{i,j}=1\}\sI\{W_{i,q}=1\} }{\prob(W_{i,j}=1)\prob(W_{i,q}=1)}{Y}_{v_{i,j},v_{i,j+1}}\cdot{Y}_{v_{i,q},v_{i,q+1}}\right]-{Y}_{v_{i,j},v_{i,j+1}}\cdot {Y}_{v_{i,q},v_{i,q+1}}\notag\\ \notag
&=\left(\frac{\prob(W_{i,j}=1,W_{i,q}=1)}{\prob(W_{i,j}=1)\prob(W_{i,q}=1)}-1\right) {Y}_{v_{i,j},v_{i,j+1}}\cdot {Y}_{v_{i,q},v_{i,q+1}}\\ \label{eq:cov-last-step}
&= \left(\frac{\prob(W_{i,q}=1 \mid W_{i,j}=1)}{\prob(W_{i,q}=1)}-1\right) {Y}_{v_{i,j},v_{i,j+1}}\cdot {Y}_{v_{i,q},v_{i,q+1}}.
\end{align}
Thus, we now only need to calculate $\prob(W_{i,q}=1 \mid W_{i,j}=1)$.
\begin{align*}
    \prob(W_{i,q}=1 \mid W_{i,j}=1)&=\prob(W_{i,q}=1, W_{i,q-1}=0 \mid W_{i,j}=1)\\
    &=\prob(W_{i,q}=1\mid W_{i,q-1}=0 ) \prob(W_{i,q-1}=0\mid W_{i,j}=1) \\
    &=p \cdot (1-\prob(W_{i,q-1}=1\mid W_{i,j}=1)).
\end{align*}
With the above equation, we can resort to recursion that
\begin{equation}\label{eq:one-condition-one}
    \prob(W_{i,q}=1 \mid W_{i,j}=1)=\frac{p}{p+1}+(-p)^{q-j}\left(\prob(W_{i,j}=1 \mid W_{i,j}=1)-\frac{p}{p+1}\right)=\frac{p+(-p)^{q-j}}{p+1}.
\end{equation}
Putting \Eqref{eq:uncondition-prob} and \Eqref{eq:one-condition-one} back to \Eqref{eq:cov-last-step}, we can have
\begin{align}\label{eq:cov-single-term}
\Cov\left(\tilde{Y}_{v_{i,j},v_{i,j+1}},\tilde{Y}_{v_{i,q},v_{i,q+1}}\right)=-(-p)^{q-j-1}{Y}_{v_{i,j},v_{i,j+1}}\cdot {Y}_{v_{i,q},v_{i,q+1}}
\end{align}
With plugging \Eqref{eq:variance-single-term} and \Eqref{eq:cov-single-term} to \Eqref{eq:variance-decompose}, we finish the proof for the path case.

Under the cycle case, for $j=1,\cdots,k(i)-1$,  $\Var(\tilde{Y}_{v_{i,j},v_{i,j+1}})$ still follows \Eqref{eq:variance-single-term}. Also, for any $1\le j<q\le k(i)-1$, the covariance $\Cov(\tilde{Y}_{v_{i,j},v_{i,j+1}},\tilde{Y}_{v_{i,q},v_{i,q+1}})$ are still of the same form in \Eqref{eq:cov-single-term}. What we need to additionally calculate is the term related to the last edge, i.e.,  $\Var(\tilde{Y}_{v_{i,k(i)},v_{i,k(i)+1}})$ and $\Cov(\tilde{Y}_{v_{i,j},v_{i,j+1}},\tilde{Y}_{v_{i,k(i)},v_{i,k(i)+1}})$ for $1\le j\le k(i)-1$. The variance part is similar as before with a different plug-in of the probability.
\begin{align}  \Var\left(\tilde{Y}_{v_{i,k(i)},v_{i,k(i)+1}}\right)
    &=\frac{1-\prob(W_{i,k(i)}=1)}{\prob(W_{i,k(i)}=1)} Y_{v_{i,k(i)},v_{i,k(i)+1}}^2\notag\\
    &=\frac{p^2+2p+(-p)^{k(i)-1}}{1-(-p)^{k(i)-1}} Y_{v_{i,k(i)},v_{i,k(i)+1}}^2.\label{eq:variance-last-edge}
\end{align}

For the covariance term, \Eqref{eq:cov-last-step} still holds and the key challenge here is to calculate $\prob(W_{i,k(i)}=1 \mid W_{i,j}=1)$ for $1\le j \le k(i)-1$. 
\begin{align*}
    \prob(W_{i,k(i)}=1 \mid W_{i,j}=1)
    &=\prob(W_{i,k(i)-1}=0, W_{i,1}=0\mid W_{i,j}=1)\\
    &=\prob(W_{i,k(i)-1}=0\mid W_{i,1}=0, W_{i,j}=1) \prob( W_{i,1}=0 \mid W_{i,j}=1)\\
    &=\prob(W_{i,k(i)-1}=0\mid W_{i,j}=1) \prob( W_{i,1}=0 \mid W_{i,j}=1)\\
    &=\prob(W_{i,k(i)-1}=0\mid W_{i,j}=1) \frac{\prob( W_{i,j}=1  \mid W_{i,1}=0)\prob( W_{i,1}=0)}{\prob( W_{i,j}=1)},
\end{align*}
where the first equation holds due to our AP design that $W_{i,k(i)}=1$ if and only if $W_{i,k(i)-1}=W_{i,1}=0$, the third equality is based on the fact that $W_{i,k(i)-1}$ is independent with $W_{i,1}$ conditioned on $W_{i,j}$, and the last equality adopts the Bayes' law. By \Eqref{eq:zero-to-zero} and \Eqref{eq:one-condition-one}, we can have
\begin{align*}
    &\prob(W_{i,k(i)-1}=0\mid W_{i,j}=1)=1- \prob(W_{i,k(i)-1}=1\mid W_{i,j}=1) = \frac{1-(-p)^{k(i)-1-j}}{p+1},\\
    &\prob(W_{i,j}=1\mid W_{i,1}=0)=1- \prob(W_{i,j}=0\mid W_{i,1}=0) = \frac{p+(-p)^{j}}{p+1}.
\end{align*}
Together with \Eqref{eq:uncondition-prob}, we can now have,
\begin{equation}\label{eq:last-edge-one-on-one}
    \prob(W_{i,k(i)}=1 \mid W_{i,j}=1)= \frac{(1-(-p)^{j-1})(1-(-p)^{k(i)-1-j})}{(1+p)^2}.
\end{equation}	
By plugging \Eqref{eq:last-edge-one-on-one} into \Eqref{eq:cov-last-step}, we can have
\begin{align*}
\Cov(\tilde{Y}_{v_{i,j},v_{i,j+1}},\tilde{Y}_{v_{i,k(i)},v_{i,k(i)+1}})=\left(\frac{(1-(-p)^{k(i)-1-j})(1-(-p)^{j-1})}{(1-(-p)^{k(i)-1})}-1\right)Y_{v_{i,j},v_{i,j+1}}Y_{v_{i,k(i)},v_{i,k(i)+1}}.
\end{align*}
We finish the proof. \qed
\end{proof}	

\section{Proof to Proposition \ref{prop:max-variance}}
\begin{proof}{Proof to Proposition \ref{prop:max-variance}}
    The core idea in this proof is to verify that all the coefficients in Theorem \ref{thm:variance-decompose} on $\path_i$ are positive.


    
    The only challenging term requiring checking is the covariance with the last edge,
    \begin{align*}
        &(-1)^j \left(\frac{(1-(-p)^{k(i)-1-j})(1-(-p)^{j-1})}{(1+p^{k(i)-1})}-1\right)\\
        &= (-1)^j \frac{-(-p)^{j-1}-(-p)^{k(i)-1-j}+(-p)^{k(i)-2}-p^{k(i)-1}}{1+p^{k(i)-1}}\\
        &= \frac{p^{j-1}+p^{k(i)-1-j}+(-1)^{j}p^{k(i)-2}+(-1)^{j+1}p^{k(i)-1}}{1+p^{k(i)-1}}\\
        &\ge 0,
    \end{align*}
    where the last step holds since $p^{j-1}\ge -p^{k(i)-1}$ and $p^{k(i)-1-j}\ge -p^{k(i)-2}$.

    Therefore, with all the positive coefficients, the maximization problem \Eqref{eq:max-variance} is equivalent to maximizing $Y_{v_{i,j},v_{i,j+1}}^2$ and $Y_{v_{i,j},v_{i,j+1}}Y_{v_{i,q},v_{i,q+1}}$ for all $i,j,q$. Thus, the optimal solution is $Y_{i,j}=B$ for all $(i,j)\in\sM$. We finish the proof. \qed
\end{proof} 

\section{Proof to Theorem \ref{thm:variance-upper-bound}} 

\begin{proof}{Proof to Theorem \ref{thm:variance-upper-bound}}
    For the alternating path $\path_s=\{v_{s,1},\cdots,v_{s,k(s)+1}\}\in\Path$, by plugging Proposition \ref{prop:max-variance} into \Eqref{eq:variance-of-path}, we can get
    \begin{align} 
    \max_{\substack{ |Y_{i,j}|\le B  \\ \forall (i,j)\in\path_s }} \Var (\EPTE_s) &= \frac{k(s)}{p} B^2 + 2 \sum_{1\le j<q \le k(s)} p^{q-j-1} B^2\notag\\
    &=\frac{k(s)}{p} B^2 + 2\sum_{1\le j < k(s)-1}p^{-j-1} B^2 \sum_{j<q\le k(s)} p^q\notag\\
    &=\frac{k(s)}{p} B^2 + 2\sum_{1\le j < k(s)-1}p^{-j-1} B^2 \frac{p^{j+1}(1-p^{k(s)-j})}{1-p}\notag\\
    &=\frac{k(s)}{p} B^2 + \frac{2B^2}{1-p}\sum_{1\le j < k(s)-1} (1-p^{k(s)-j})\notag\\
    &=\frac{k(s)}{p} B^2 + \frac{2B^2}{1-p} \left(k(s)-1-\frac{p^{k(s)}-p}{p-1}\right).
    \label{eq:app-variance-upper-1}
\end{align}

\end{proof}

\section{Proof to Theorem \ref{thm:clt}}
First, we need to formally need the triangular array since for each given $N$, the matching can vary dramatically. Let us denote $Z_N:=N(\AP-\ATE)$ and we have
\begin{equation}
    Z_N=N(\AP-\ATE)=\sum_{\path_{i,N} \in \Path_N} \EPTE_{i,N}- \PTE_{i,N},
\end{equation}
where the new subscript $N$ is the index and recall that 
\begin{align*}
   &\EPTE_{i,N}= \sum_{\substack{ (v_{i,j},v_{i,j+1}) \in \\  \path_{i,N} \cap \Diff_{t,N}}}\frac{\sI\{w_{i,j}=1\}\cdot Y_{v_{i,j},v_{i,j+1}}}{\prob(W_{i,j}=1)}-\sum_{ \substack{ (v_{i,j},v_{i,j+1}) \in \\  \path_{i,N} \cap \Diff_{c,N}}}\frac{\sI\{w_{i,j}=1\}\cdot Y_{v_{i,j},v_{i,j+1}}}{\prob(W_{i,j}=1)},\\
   &\PTE_{i,N}=\sum_{\substack{ (v_{i,j},v_{i,j+1}) \in \\  \path_{i,N} \cap \Diff_{t,N}}} Y_{v_{i,j},v_{i,j+1}}-\sum_{\substack{ (v_{i,j},v_{i,j+1}) \in \\  \path_{i,N} \cap \Diff_{c,N}}} Y_{v_{i,j},v_{i,j+1}}.
\end{align*}

We now partition $\Path_N$ into giant sequences  $\gG_N$ and small sequences $\gS_N$, where
\begin{align*}
    &\gG_N = \{\path_{i,N} \in \Path_N: \Var(\EPTE_{i,N}) \ge (\Var(Z_N))^{\frac{2}{3}} \},\\
    &\gS_N = \{\path_{i,N} \in \Path_N: \Var(\EPTE_{i,N}) < (\Var(Z_N))^{\frac{2}{3}} \}.
\end{align*}
For simplicity, we introduce the notation that $Z_{i,N}=\EPTE_{i,N}- \PTE_{i,N}$, $Z_{G,N}=\sum_{\path_{i,N} \in \gG_N} Z_{i,N}$ and $Z_{S,N}=\sum_{\path_{i,N} \in \gS_N} Z_{i,N}$. Then, we have $Z_N=Z_{S,N}+Z_{L,N}$, and by independence across all the paths, $\Var(Z_N)=\Var(Z_{S,N})+\Var(Z_{L,N})$. In addition, since $\E[\AP]=\tau$, we can have $\Var(Z_N)=\Var(N\AP)$. Also, we can know that $|\gG_N |\le (\Var(Z_N))^{\frac{1}{3}}$. 

Now, we are going to show that if $\Var(Z_{G,N})/\Var(Z_N)\rightarrow \alpha$ (or $\Var(Z_{S,N})/\Var(Z_N)\rightarrow \alpha$) for any $\alpha>0$, $Z_{G,N}/\sqrt{\Var(Z_{G,N})}$ (correspondingly, $Z_{G,N}/\sqrt{\Var(Z_{G,N})}$) converges to the standard normal distribution. Note that this is still not enough, since $\Var(Z_{G,N})/\Var(Z_N)$ or $\Var(Z_{S,N})/\Var(Z_N)$ may not exist at all.

\textbf{Case 1:} If $\Var(Z_{S,N})/\Var(Z_N)\rightarrow \alpha$ for some $\alpha>0$. Let us consider the $\path_{i,N} \in \gS_N$, and we denote $\path_{i,N}:=\{v_{i,N,1},\cdots,v_{i,N,k(i,N)}, v_{i,N,k(i,N)+1}\}$. WLOG, we assume $(v_{i,N,1},v_{i,N,2})\in\Diff_{t,N}$. With this notation, we can have
\begin{equation*}
    Z_{i,N}=\sum_{j=1}^{k(i,N)}(-1)^{j+1}\frac{Y_{v_{i,j},v_{i,j+1}}}{\prob(W_{i,j}=1)}\left({\sI\{w_{i,j}=1\}}-\prob(W_{i,j}=1)\right).
\end{equation*}
From \Eqref{eq:uncondition-prob}, we know that $\prob (W_{i,j}=1)\ge \min\{p(1-p),(1-p)^2\}$. Together with our assumption of the bounded $Y_{v_{i,j},v_{i,j+1}}$, we can upper bound the absolute value of $\widetilde{Y}_{v_{i,j},v_{i,j+1}}:=(-1)^{j+1}\frac{Y_{v_{i,j},v_{i,j+1}}}{\prob(W_{i,j}=1)}$ by $\frac{B}{\min\{p(1-p),(1-p)^2\}}$. For simplicity, we introduce the positive constant $C_{p,0}$ to denote that $|\widetilde{Y}_{v_{i,j},v_{i,j+1}}|\le C_{p,0}$ uniformly. We also adopt $\widetilde{W}_{i,j}$ to denote ${\sI\{w_{i,j}=1\}}-\prob(W_{i,j}=1)$, and thus $Z_{i,N}=\sum_{j=1}^{k(i,N)} \widetilde{Y}_{v_{i,j},v_{i,j+1}} \widetilde{W}_{i,j}$. We purposely split out the last edge of $\path_{i,N}$ because if $\path_i$ is a cycle, the last edge $W_{i,k(i,N)}$ does not satisfy the Markov property, and we define $Z_{i,N}^{(-1)}:=\sum_{j=1}^{k(i,N)-1} \widetilde{Y}_{v_{i,j},v_{i,j+1}} \widetilde{W}_{i,j}$. 

Now we are going to verify the Lindeberg condition. We start with bounding the fourth moment of $Z_{i,N}$, i.e., $\E[Z_{i,N}^4]$. First, we still separate out the last edge out because $W_{i,1},\cdots,W_{i,k(i,N)-1}$ is always a Markov chain with the transition matrix:
\[
\mP = \begin{pmatrix}
1 - p & p \\
1 & 0
\end{pmatrix},
\]
the eigenvalues of $\mP$ are $1$ and $-p$. The absolute spectral gap $\gamma^*$ is $1-|-p|=1-p$. We can have the fast mixing properties. To be more specific, we have
\begin{align}
    \E[Z_{i,N}^4] &\le \E\left[\left(Z_{i,N}^{(-1)}+\widetilde{Y}_{v_{i,k(i,N)},v_{i,k(i,N)+1}}\widetilde{W}_{i,k(i,N)}\right)^4\right]\notag\\
    &\le 8 \E\left[\left(Z_{i,N}^{(-1)}\right)^4\right]+8\E\left[\left(\widetilde{Y}_{v_{i,k(i,N)},v_{i,k(i,N)+1}\widetilde{W}_{i,k(i,N)}}\right)^4\right]\notag\\
    &\le 8 \E\left[\left(Z_{i,N}^{(-1)}\right)^4\right]+8\widetilde{Y}_{v_{i,k(i,N)},v_{i,k(i,N)+1}}^4\label{eq:fourth-moment-decompose},
\end{align}
where the inequality is based on the Cauthy-Schwarz inequality. For the first term, we are applying the Rosenthal-type inequality for the strong mixing sequence by \cite{rio2017asymptotic} (see, also in Theorem \ref{thm:fourth-moment}). Define $Q_j(u)$ as the quantile function of $\widetilde{Y}_{v_{i,j},v_{i,j+1}}\widetilde{W}_{i,j}$ for $u\in(0,1)$. Since $|\widetilde{W}_{i,j}|\le 1$, $Q_k(u)\le|\widetilde{Y}_{v_{i,j},v_{i,j+1}}|$. As the chain is geometrically mixing, the strong mixing coefficients $\alpha(m)\le c_\alpha \rho^m$ with $\rho\in(0,1)$. Thus, there exists positive constants $a_{p}$ and $b_{p}$ only dependent on $p$ such that, for all $u\in(0,1)$,
\begin{equation*}
    \alpha^{-1}(u)\le a_p+b_p\log\left(\frac{1}{u}\right),
\end{equation*}
where $\alpha^{-1}(u):=\inf\{m\ge 0: \alpha(m)\le u\}$. Following the notation in Theorem \ref{thm:fourth-moment},
\begin{align}
    M_{4,\alpha,k(i,N)-1}(Q_j)&=\sum_{j=1}^{k(i,N)-1}\int_0^1[\alpha^{-1}(u)\land n]^3 Q_k^4(u)du\notag \\
    &\le \sum_{j=1}^{k(i,N)-1}\int_0^1[\alpha^{-1}(u)]^3 (\widetilde{Y}_{v_{i,j},v_{i,j+1}})^4du\notag\\
    &\le \sum_{j=1}^{k(i,N)-1} (\widetilde{Y}_{v_{i,j},v_{i,j+1}})^4 \int_0^{1} (a_p-b_p\log(u))^3 du\notag\\
    &= (a_p^3+3a_p^2b_p+6a_p b_p^2+6 b_p^3) \sum_{j=1}^{k(i,N)-1} (\widetilde{Y}_{v_{i,j},v_{i,j+1}})^4\notag\\
    &:=C_\alpha(p) \sum_{j=1}^{k(i,N)-1} (\widetilde{Y}_{v_{i,j},v_{i,j+1}})^4.\label{eq:M-upper-bound}
\end{align}

For any $1\le \gamma\le\kappa \le k(i,N)-1$, we have
\begin{align*}
    \E\left[(\widetilde{Y}_{v_{i,\gamma},v_{i,\gamma+1}}\widetilde{W}_{i,\gamma})(\widetilde{Y}_{v_{i,\kappa},v_{i,\kappa+1}}\widetilde{W}_{i,\kappa})\right]&=\widetilde{Y}_{v_{i,\gamma},v_{i,\gamma+1}} \widetilde{Y}_{v_{i,\kappa},v_{i,\kappa+1}} \E[\widetilde{W}_{i,\gamma}\widetilde{W}_{i,\kappa}]\\
    &= \widetilde{Y}_{v_{i,\gamma},v_{i,\gamma+1}} \widetilde{Y}_{v_{i,\kappa},v_{i,\kappa+1}} \sigma_W^2 (-p)^{\kappa-\gamma},
\end{align*}
where $\sigma_W^2=p/(1+p)^2$. With this, we can further have,
\begin{align}
    \sum_{\kappa=1}^{k(i,N)-1}\sum_{\gamma=1}^{k(i,N)-1} \left|\E\left[(\widetilde{Y}_{v_{i,\gamma},v_{i,\gamma+1}}\widetilde{W}_{i,\gamma})(\widetilde{Y}_{v_{i,\kappa},v_{i,\kappa+1}}\widetilde{W}_{i,\kappa})\right]\right|&=\sigma_W^2 \sum_{\kappa=1}^{k(i,N)-1}\sum_{\gamma=1}^{k(i,N)-1}|\widetilde{Y}_{v_{i,\gamma},v_{i,\gamma+1}} \widetilde{Y}_{v_{i,\kappa},v_{i,\kappa+1}}|  p^{|\kappa-\gamma|}\notag\\
    &\le \sigma_W^2\sum_{\kappa=1}^{k(i,N)-1}\sum_{\gamma=1}^{k(i,N)-1}\frac{\widetilde{Y}_{v_{i,\gamma},v_{i,\gamma+1}}^2+ \widetilde{Y}_{v_{i,\kappa},v_{i,\kappa+1}}^2}{2}  p^{|\kappa-\gamma|}\notag\\
    &\le \sigma_W^2 \frac{1+p}{1-p} \sum_{j=1}^{k(i,N)-1}\widetilde{Y}_{v_{i,j},v_{i,j+1}}^2. \label{eq:cov-upper-bound}
\end{align}
Applying \Eqref{eq:M-upper-bound} and \Eqref{eq:cov-upper-bound} to Theorem \ref{thm:fourth-moment}, we can have
\begin{align}
    \E\left[\left(Z_{i,N}^{(-1)}\right)^4\right]&\le 3\left(\sigma_W^2 \frac{1+p}{1-p} \sum_{j=1}^{k(i,N)-1}\widetilde{Y}_{v_{i,j},v_{i,j+1}}^2\right)^2 + 48 C_\alpha(p) \sum_{j=1}^{k(i,N)-1} \widetilde{Y}_{v_{i,j},v_{i,j+1}}^4,\notag\\
    &=3\left(\sigma_W^2 \frac{1+p}{1-p} \sum_{j=1}^{k(i,N)-1}\widetilde{Y}_{v_{i,j},v_{i,j+1}}^2\right)^2 + 48 C_\alpha(p) \sum_{j=1}^{k(i,N)-1} \widetilde{Y}_{v_{i,j},v_{i,j+1}}^4\notag\\
    &=3\left( \frac{1+p}{1-p} \sum_{j=1}^{k(i,N)-1}\Var(\widetilde{Y}_{v_{i,j},v_{i,j+1}}\widetilde{W}_{i,j})\right)^2 + 48 C_\alpha(p) \sum_{j=1}^{k(i,N)-1} \widetilde{Y}_{v_{i,j},v_{i,j+1}}^4\notag\\
    &\le 3\left( \frac{1+p}{1-p} \sum_{j=1}^{k(i,N)}\Var(\widetilde{Y}_{v_{i,j},v_{i,j+1}}\widetilde{W}_{i,j})\right)^2 + 48 C_\alpha(p) \sum_{j=1}^{k(i,N)-1} \widetilde{Y}_{v_{i,j},v_{i,j+1}}^4\notag\\
    &\le 3\left( \frac{1+p}{1-p} \Var(Z_{i,N})\right)^2 + 48 C_\alpha(p) \sum_{j=1}^{k(i,N)-1} \widetilde{Y}_{v_{i,j},v_{i,j+1}}^4,\label{eq:Z-1-4-bound}
\end{align}
where the last inequality again adopts the fact that all the correlation terms are positive. Putting \Eqref{eq:Z-1-4-bound} back to \Eqref{eq:fourth-moment-decompose}, we get
\begin{align*}
    \E[Z_{i,N}^4]&\le 24\left( \frac{1+p}{1-p} \Var(Z_{i,N})\right)^2 + 384 C_\alpha(p) \sum_{j=1}^{k(i,N)-1} \widetilde{Y}_{v_{i,j},v_{i,j+1}}^4+8\widetilde{Y}_{v_{i,k(i,N)},v_{i,k(i,N)+1}}^4\\
    &\le  24\left( \frac{1+p}{1-p} \Var(Z_{i,N})\right)^2 + C_{p,2} \sum_{j=1}^{k(i,N)} \widetilde{Y}_{v_{i,j},v_{i,j+1}}^4\\
    &\le 24\left( \frac{1+p}{1-p} \Var(Z_{i,N})\right)^2 + C_{p,2} C_{p,0}^2\sum_{j=1}^{k(i,N)} \widetilde{Y}_{v_{i,j},v_{i,j+1}}^2\\
    &\le 24\left( \frac{1+p}{1-p} \Var(Z_{i,N})\right)^2 + \frac{C_{p,2} C_{p,0}^2}{\sigma_W^2}\sum_{j=1}^{k(i,N)} \Var(\widetilde{Y}_{v_{i,j},v_{i,j+1}}\widetilde{W}_{i,j})\\
    &\le 24\left( \frac{1+p}{1-p} \Var(Z_{i,N})\right)^2 + \frac{C_{p,2} C_{p,0}^2}{\sigma_W^2}\Var(Z_{i,N})\\
    &:= C_{p,3} (\Var(Z_{i,N}))^2+C_{p,4}  \Var(Z_{i,N}),
\end{align*}
where $C_{p,2}:=(384 C_\alpha(p))\vee 8$. Therefore, for the set $\gS_N$ under the condition that $\Var(Z_{S,N})/\Var(Z_N)\rightarrow \alpha$, we have for any $\epsilon>0$,
\begin{align*}
     &\frac{1}{Var(Z_{S,N})}\sum_{\path_{i,N}\in \gS_N} \E\left[Z_{i,N}^2\sI\left\{|Z_{i,N}|>\epsilon \sqrt{\Var(Z_{S,N})}\right\}\right] \\
     &\qquad\le  \frac{1}{Var(Z_{S,N})}\sum_{\path_{i,N}\in \gS_N} \frac{\E\left[Z_{i,N}^4\right]}{\epsilon^2 \Var(Z_{S,N})}\\
     &\qquad \le \frac{1}{\epsilon^2(Var(Z_{S,N}))^2} \sum_{\path_{i,N}\in \gS_N} C_{p,3} (\Var(Z_{i,N}))^2+C_{p,4}  \Var(Z_{i,N})\\
     &\qquad \le \frac{1}{\epsilon^2(Var(Z_{S,N}))^2} \sum_{\path_{i,N}\in \gS_N} C_{p,3} (\Var(Z_N))^{\frac{2}{3}} \Var(Z_{i,N})+C_{p,4}  \Var(Z_{i,N})\\
     &\qquad = \frac{C_{p,3} (\Var(Z_N))^{\frac{2}{3}}+C_{p,4}}{\epsilon^2(Var(Z_{S,N}))^2} \sum_{\path_{i,N}\in \gS_N} \Var(Z_{i,N})\\
     &\qquad = \frac{C_{p,3} (\Var(Z_N))^{\frac{2}{3}}+C_{p,4}}{\epsilon^2(Var(Z_{S,N}))} \\
     &\qquad\rightarrow 0,
\end{align*}
where the first inequality holds due to the fact that $\E[X^2 \sI\{|X|>s\}]\le \E[X^4]/s^2$, the third inequality is because of the definition of $\gS_N$, the second equality is based on the indepedence among all the paths, and the last step is because $\Var(Z_{S,N})/\Var(Z_N)\rightarrow \alpha$ and $\Var(Z_N)\rightarrow\infty$. Therefore, we can conclude that if the limit of $\Var(Z_{S,N})/\Var(Z_N)$ exists, we can guarantee that as $N\rightarrow\infty$
\begin{equation}\label{eq:case-1-CLT}
    \frac{\sum_{\path_{i,N}\in \gS_N}Z_{i,N}}{\sqrt{\Var(Z_{S,N})}}\xrightarrow{d} \gN(0,1).
\end{equation}

\textbf{Case 2:} If $\Var(Z_{G,N})/\Var(Z_N)\rightarrow \beta$ for some $\beta>0$. Let us consider the $\path_{i,N} \in \gG_N$, and we denote $\path_{i,N}:=\{v_{i,N,1},\cdots,v_{i,N,k(i,N)}, v_{i,N,k(i,N)+1}\}$. WLOG, we assume $(v_{i,N,1},v_{i,N,2})\in\Diff_{t,N}$. Following the same notation in Case 1, we can have 
\begin{equation*}
    Z_{i,N}= Z_{i,N}^{(-1)}+\widetilde{Y}_{v_{i,k(i,N)},v_{i,k(i,N)+1}}\widetilde{W}_{i,j}.
\end{equation*}
Then, we can get
\begin{equation}\label{eq:giant-variance}
    \frac{\sum_{\path_{i,N}\in \gG_N}Z_{i,N}}{\sqrt{\Var(Z_{G,N})}} = \frac{\sum_{\path_{i,N}\in \gG_N}Z_{i,N}^{(-1)}}{\sqrt{\Var(Z_{G,N})}}+\frac{\sum_{\path_{i,N}\in \gG_N}\widetilde{Y}_{v_{i,k(i,N)},v_{i,k(i,N)+1}}\widetilde{W}_{i,j}}{\sqrt{\Var(Z_{G,N})}}.
\end{equation}
Let us first focus on the second term of \Eqref{eq:giant-variance}. Based on the definition of $\gG_N$, we can guarantee that $|\gG_N|\le (\Var(Z_N))^{\frac{1}{3}}$. As the assignment across different paths is independent from each other, 
\begin{align}
    \Var\left(\sum_{\path_{i,N}\in \gG_N}\widetilde{Y}_{v_{i,k(i,N)},v_{i,k(i,N)+1}}\widetilde{W}_{i,j}\right) &= \sum_{\path_{i,N}\in \gG_N} \Var\left(\widetilde{Y}_{v_{i,k(i,N)},v_{i,k(i,N)+1}}\widetilde{W}_{i,j}\right)\notag\\
    &\le \sum_{\path_{i,N}\in \gG_N}  C_{p,0}^2 \Var(\widetilde{W}_{i,j})\notag\\
    &\le \frac{C_{p,0}^2}{4}(\Var(Z_N))^{\frac{1}{3}},\label{eq:tail-variance-bound}
\end{align}
where the last inequality holds due to the facts that $\Var(\widetilde{W}_{i,j})\le \frac{1}{4}$ and $|\gG_N|\le (\Var(Z_N))^{\frac{1}{3}}$. Therefore, we can have that the second term of \Eqref{eq:giant-variance} will converge to zero as $\Var(Z_{G,N})/\Var(Z_N)\rightarrow \beta$, i.e.,
\begin{equation}
    \lim_{N\rightarrow\infty} \frac{\sum_{\path_{i,N}\in \gG_N}\widetilde{Y}_{v_{i,k(i,N)},v_{i,k(i,N)+1}}\widetilde{W}_{i,k(i,N)}}{\sqrt{\Var(Z_{G,N})}} \le  \lim_{N\rightarrow\infty} \frac{\frac{C_{p,0}^2}{4}(\Var(Z_N))^{\frac{1}{3}}}{\sqrt{\Var(Z_{G,N})}}=0.\label{eq:last-edge-small}
\end{equation}

For the first term in \Eqref{eq:giant-variance}, we are applying the non-stationary $\alpha$-mixing CLT by \cite{bradley2017central} (see, also Theorem \ref{thm:non-stationary-CLT}). For $\path_{i,N}$, we define $\gC_{i,N}^{(-1)}$  as $W_{i,1},\cdots,W_{i,k(i,N)-1}$, which is a Markov chain. We then concatenate the $\gC_{i,N}^{(-1)}$ for every $\path_{i,N}$ in $\gG_N$ together and get a long chain $\gC_N$. For the chain $\gC_N$, in general, it is not a Markov Chain anymore since between the paths, they are independence. However, the independence still preserves an $\alpha$-mixing sequence. The maximal correlation coefficient just decreases when the two $\sigma$-fields are cross the paths. In order to make the Lindeberg condition hold, we now show that $\Var(\sum_{\path_{i,N}\in \gG_N}Z_{i,N}^{(-1)})$ goes to infinity. Consider that
\begin{align}
    &\lim_{N\rightarrow\infty}\frac{\Var(\sum_{\path_{i,N}\in \gG_N}Z_{i,N}^{(-1)})}{\Var(Z_{G,N})}=\lim_{N\rightarrow\infty}\frac{\Var(Z_{G,N}-\sum_{\path_{i,N}\in \gG_N}\widetilde{Y}_{v_{i,k(i,N)},v_{i,k(i,N)+1}}\widetilde{W}_{i,k(i,N)})}{\Var(Z_{G,N})}\notag\\
    &=1+\lim_{N\rightarrow\infty}\frac{\Var(\sum_{\path_{i,N}\in \gG_N}\widetilde{Y}_{v_{i,k(i,N)},v_{i,k(i,N)+1}}\widetilde{W}_{i,k(i,N)})-2\Cov(\sum_{\path_{i,N}\in \gG_N}\widetilde{Y}_{v_{i,k(i,N)},v_{i,k(i,N)+1}}\widetilde{W}_{i,k(i,N)},Z_{G,N})}{\Var(Z_{G,N})}\notag\\
    &=1-2\lim_{N\rightarrow\infty}\frac{\Cov(\sum_{\path_{i,N}\in \gG_N}\widetilde{Y}_{v_{i,k(i,N)},v_{i,k(i,N)+1}}\widetilde{W}_{i,k(i,N)},Z_{G,N})}{\Var(Z_{G,N})}\notag\\
    &=1, \label{eq:variance-variance-wo-last-edge}
\end{align}
where the third equality holds due to \Eqref{eq:tail-variance-bound} and $\Var(Z_{G,N})/\Var(Z_N)\rightarrow \beta$ and the last equality is based on the observation that
\begin{align*}
    \left|\Cov(\sum_{\path_{i,N}\in \gG_N}\widetilde{Y}_{v_{i,k(i,N)},v_{i,k(i,N)+1}}\widetilde{W}_{i,k(i,N)},Z_{G,N})\right| &\le \sqrt{\Var(\sum_{\path_{i,N}\in \gG_N}\widetilde{Y}_{v_{i,k(i,N)},v_{i,k(i,N)+1}}\widetilde{W}_{i,k(i,N)})\Var(Z_{G,N})}\\
    &\le \sqrt{\frac{C_{p,0}^2}{4}(\Var(Z_N))^{\frac{1}{3}}\Var(Z_{G,N})},
\end{align*}
and then the limit goes to zero. Thus, \Eqref{eq:variance-variance-wo-last-edge} indicates that $\Var(\sum_{\path_{i,N}\in \gG_N}Z_{i,N}^{(-1)})$ goes to infinity. Moreover, since $\widetilde{Y}_{v_{i,j},v_{i,j+1}}\widetilde{W}_{i,j}$ is bounded by a constant, the Lindeberg condition easily holds. Therefore, we know that 
\begin{equation*}
    \frac{\sum_{\path_{i,N}\in \gG_N}Z_{i,N}^{(-1)}}{\sqrt{\Var(\sum_{\path_{i,N}\in \gG_N}Z_{i,N}^{(-1)})}}\xrightarrow{d} \gN(0,1).
\end{equation*}
By \Eqref{eq:last-edge-small}, \Eqref{eq:variance-variance-wo-last-edge}, and Slutsky's theorem, we can have that as $N\rightarrow\infty$,
\begin{equation}\label{eq:case-2-CLT}
    \frac{\sum_{\path_{i,N}\in \gG_N}Z_{i,N}}{\sqrt{\Var(Z_{G,N})}}\xrightarrow{d} \gN(0,1).
\end{equation}

Now, we are finalizing the proof by establishing the result for the case when $\alpha$ and $\beta$ may not exist. Define the following notations,
\begin{equation*}
    \alpha_N = \frac{\Var(Z_{S,N})}{\Var(Z_N)},\qquad \beta_N=\frac{\Var(Z_{G,N})}{\Var(Z_{N})}.
\end{equation*}
It is natural to see $\alpha_N+\beta_N=1$ and both of them are in $[0,1]$. Thus, $(\alpha_N,\beta_N)$ always lies in a compact set. By the Bolzano-Weierstrass theorem, for every subsequence $(\alpha_{N_k},\beta_{N_k})$ of  $(\alpha_N,\beta_N)$, there exists a further subsubsequence $(\alpha_{N_{k_j}},\beta_{N_{k_j}})$ converging to some $(\alpha,\beta)$ with $\alpha+\beta=1$. Therefore, by Case 1 (\Eqref{eq:case-1-CLT}) and Case 2 (\Eqref{eq:case-2-CLT}), we can establish, along the subsubsequence $(\alpha_{N_{k_j}},\beta_{N_{k_j}})$, 
\begin{equation*}
    \left(\frac{\sum_{\path_{i,N_{k_j}}\in \gS_{N_{k_j}}}Z_{i,N_{k_j}}}{\sqrt{\Var(Z_{S,N_{k_j}})}},\frac{\sum_{\path_{i,N_{k_j}}\in \gG_{N_{k_j}}}Z_{i,N_{k_j}}}{\sqrt{\Var(Z_{G,N_{k_j}})}}\right) \xrightarrow{d} (Z_1,Z_2),
\end{equation*}
where $Z_1$ and $Z_2$ are i.i.d. normal. By the continuous mapping theorem,
\begin{equation*}
    \sqrt{\alpha_{N_{k_j}}}\frac{\sum_{\path_{i,N_{k_j}}\in \gS_{N_{k_j}}}Z_{i,N_{k_j}}}{\sqrt{\Var(Z_{S,N_{k_j}})}}+\sqrt{\beta_{N_{k_j}}}\frac{\sum_{\path_{i,N_{k_j}}\in \gG_{N_{k_j}}}Z_{i,N_{k_j}}}{\sqrt{\Var(Z_{G,N_{k_j}})}}\xrightarrow{d}\sqrt{\alpha_{N_{k_j}}} Z_1+\sqrt{\beta_{N_{k_j}}} Z_2,
\end{equation*}
and we know that $\sqrt{\alpha_{N_{k_j}}} Z_1+\sqrt{\beta_{N_{k_j}}} Z_2 \sim \gN(0,\alpha_{N_{k_j}}+\beta_{N_{k_j}})$ which is $\gN(0,1)$. Thus, we are same to draw the conclusion that for any subsequence of $(\alpha_N,\beta_N)$, we can always have a further subsubsequence, along which our estimator converge to the same distribution $\gN(0,1)$. By the subsequence principle (see, Theorem 2.6. in \citealt{billingsley2013convergence}), we can guarantee that if $\Var(Z_N)$ goes to infinity,
\begin{equation*}
    \frac{Z_N}{\sqrt{\Var(Z_N)}} \xrightarrow{d} \gN(0,1).
\end{equation*}
We finish the proof. \qed

\section{Technical Details to Proposition \ref{prop:estimated-variance}} \label{app:sec:var-upper-bound}

\begin{proposition}
    For $\path_i=\{v_{i,1},\cdots,v_{i,k(i)+1}\}\in\Path$, if $\path_i$ is a path, we have the following upper bound $\tilde{\sigma}^2_i$ of $\Var(\EPTE_i)$,
\begin{equation}
    \tilde{\sigma}^2_i := \left(\frac{1}{p}+1\right) \sum_{j=1}^{k(i)} Y_{v_{i,j},v_{i,j+1}}^2+\sum_{j=2}^{k(i)-1} Y_{v_{i,j},v_{i,j+1}}^2 + 2 \sum_{\substack{1\le j \le k(i)-2 \\ j+2\le q \le k(i)}} p^{q-j-1} Y_{v_{i,j},v_{i,j+1}}Y_{v_{i,q},v_{i,q+1}}.
\end{equation}
Moreover, $\tilde{\sigma}^2_i$ can be estimated by
\begin{align*}
    \hat{\sigma}^2_i&:= \left(\frac{1}{p}+1\right) \sum_{j=1}^{k(i)} \frac{\sI\{w_{i,j}=1\}\cdot Y^2_{v_{i,j},v_{i,j+1}}}{\prob(W_{i,j}=1)}+\sum_{j=2}^{k(i)-1} \frac{\sI\{w_{i,j}=1\}\cdot Y^2_{v_{i,j},v_{i,j+1}}}{\prob(W_{i,j}=1)}\\
    &\qquad\qquad\qquad + 2 \sum_{\substack{1\le j \le k(i)-2 \\ j+2\le q \le k(i)}} p^{q-j-1} \frac{Y_{v_{i,j},v_{i,j+1}}Y_{v_{i,q},v_{i,q+1}}\sI\{w_{i,j}=1,w_{i,q}=1\}}{\prob(W_{i,j}=1,W_{i,q}=1)},
\end{align*}
where $\prob(W_{i,j}=1)=p/(p+1)$ and $\prob(W_{i,j}=1,W_{i,q}=1)=(p^2-(-p)^{q-j+1})/(p+1)^2$. 

Similarly, if $\path_i$ is a cycle, the upper bound $\tilde{\sigma}^2_i$ of $\Var(\EPTE_i)$ is 
\begin{align*}
    \tilde{\sigma}^2_i :=& \left(\frac{1}{p}+2\right) \sum_{j=1}^{k(i)-1} Y_{v_{i,j},v_{i,j+1}}^2+ 2 \sum_{\substack{1\le j \le k(i)-3 \\ j+2\le q \le k(i)-1}} p^{q-j-1} Y_{v_{i,j},v_{i,j+1}}Y_{v_{i,q},v_{i,q+1}}\\
    &+ \left(\frac{p^2+2p-p^{k(i)-1}}{1+p^{k(i)-1}}+2\right) Y_{v_{i,k(i)},v_{i,k(i)+1}}^2 \\
    &+ 2 \sum_{2\le j \le k(i)-2} (-1)^j \left(\frac{(1-(-p)^{k(i)-1-j})(1-(-p)^{j-1})}{(1+p^{k(i)-1})}-1\right)Y_{v_{i,j},v_{i,j+1}}Y_{v_{i,k(i)},v_{i,k(i)+1}}.
\end{align*}
We can estimate the upper bound by
\begin{align*}
    \hat{\sigma}^2_i&:= \left(\frac{1}{p}+2\right) \sum_{j=1}^{k(i)-1} \frac{\sI\{w_{i,j}=1\}\cdot Y^2_{v_{i,j},v_{i,j+1}}}{\prob(W_{i,j}=1)} + 2 \sum_{\substack{1\le j \le k(i)-3 \\ j+2\le q \le k(i)-1}} p^{q-j-1} \frac{Y_{v_{i,j},v_{i,j+1}}Y_{v_{i,q},v_{i,q+1}}\sI\{w_{i,j}=1,w_{i,q}=1\}}{\prob(W_{i,j}=1,W_{i,q}=1)}\\
    &+\left(\frac{p^2+2p-p^{k(i)-1}}{1+p^{k(i)-1}}+2\right) \frac{\sI\{w_{i,k(i)}=1\} Y_{v_{i,k(i)},v_{i,k(i)+1}}^2}{\prob(W_{i,k(i)}=1)}\\
    &+ 2 \sum_{2\le j \le k(i)-2} (-1)^j \left(\frac{(1-(-p)^{k(i)-1-j})(1-(-p)^{j-1})}{(1+p^{k(i)-1})}-1\right)\frac{\sI\{w_{i,j}=w_{i,k(i)}=1\}Y_{v_{i,j},v_{i,j+1}}Y_{v_{i,k(i)},v_{i,k(i)+1}}}{\prob(W_{i,j}=1,W_{i,k(i)}=1)},
\end{align*}
where $\prob(W_{i,j}=1)=p/(p+1)$ for $1\le j \le k(i)-1$, $\prob(W_{i,k(i)}=1)=(1-(-p)^{k(i)-1})/(1+p)^2$, $\prob(W_{i,j}=1,W_{i,q}=1)=(p^2-(-p)^{q-j+1})/(p+1)^2$ for $q\not= k(i)$ and $\prob(W_{i,j}=1,W_{i,k(i)}=1)=\frac{p(1-(-p)^{j-1})(1-(-p)^{k(i)-1-j})}{(1+p)^3}$. 

Furthermore, the estimator $\hat{\sigma}^2_i$ is an unbiased estimator of $\tilde{\sigma}^2_i$, i.e., $\E[\hat{\sigma}^2_i]=\tilde{\sigma}^2_i$.
\end{proposition}

\end{document}